\documentclass{natureprintstyle}
\usepackage{graphicx}
\usepackage{amssymb,hyperref}
\usepackage[normalem]{ulem}

\usepackage{wrapfig}
\usepackage[square,sort,comma,numbers]{natbib}
\usepackage{epsfig}
\usepackage{sidecap}
\usepackage{color}
\usepackage{graphicx}

\def\Msun{${\rm M}_\odot$}

\title{New Regimes in the Observation of Core-Collapse Supernovae}

\author{Maryam Modjaz$^{1,2}$,
Claudia P. Guti\'errez$^{3}$,
and Iair Arcavi$^{4}$
}

\date{}

\begin{document}

\maketitle

\begin{affiliations}
 \item Center for Cosmology and Particle Physics, New York University, 726 Broadway, New York, NY 10003, USA \\
 \item Center for Computational Astrophysics, Flatiron Institute,162 5th Avenue, 10010, New York, NY, USA \\
 \item Department of Physics and Astronomy, University of Southampton, Southampton, SO17 1BJ, UK \\
 \item The School of Physics and Astronomy, Tel Aviv University, Tel Aviv 69978, Israel
\end{affiliations}

\begin{abstract}
Core-collapse Supernovae (CCSNe) mark the deaths of stars more massive than about eight times the mass of the sun (\Msun) and are intrinsically the most common kind of catastrophic cosmic explosions. They can teach us about many important physical processes, such as nucleosynthesis and stellar evolution,
and thus, they have been studied extensively for decades. However, many crucial questions remain unanswered, including the most basic ones regarding which kinds of massive stars achieve which kind of explosions and how. Observationally, this question is related to the open puzzles of whether CCSNe can be divided into distinct types or whether they are drawn from a population with a continuous set of properties, and of what progenitor characteristics drive the diversity of observed explosions. Recent developments in wide-field surveys and rapid-response followup facilities are helping us answer these questions by providing two new tools: (1) large statistical samples which enable population studies of the most common SNe, and reveal rare (but extremely informative) events that question our standard understanding of the explosion physics involved, and (2) observations of early SNe emission taken shortly after explosion which carries signatures of the progenitor structure and mass loss history.
Future facilities will increase our capabilities and allow us to answer many open questions related to these extremely energetic phenomena of the Universe. 
\end{abstract}

We focus this short review on a few open questions which are being tackled by new facilities for quick discovery and rapid-response followup, as well as for studying CCSNe in large numbers. For a more exhaustive review of the field see recent compilations such as the Handbook of Supernovae \citep{SNHandbook}.

\section*{The CCSN Classification Landscape}
~\\

SN classification has been a challenge since the 1940's \citep{Minkowski41}, but especially recently, as innovative surveys have brought a large increase of new and debated types. Classification schemes are important as physically motivated ones can give crucial insight into the explosion physics and stellar evolution pathways that lead to the different kinds of important explosions.

The classical classification scheme \citep[e.g.,][]{Filippenko97,Gal-Yam16} relies mostly on spectra and has two main types: Type I SNe (SNe I) which do not show hydrogen lines, and Type II SNe (SNe II) which do.

\begin{figure*}[ht!]
\includegraphics[width=.97\textwidth]{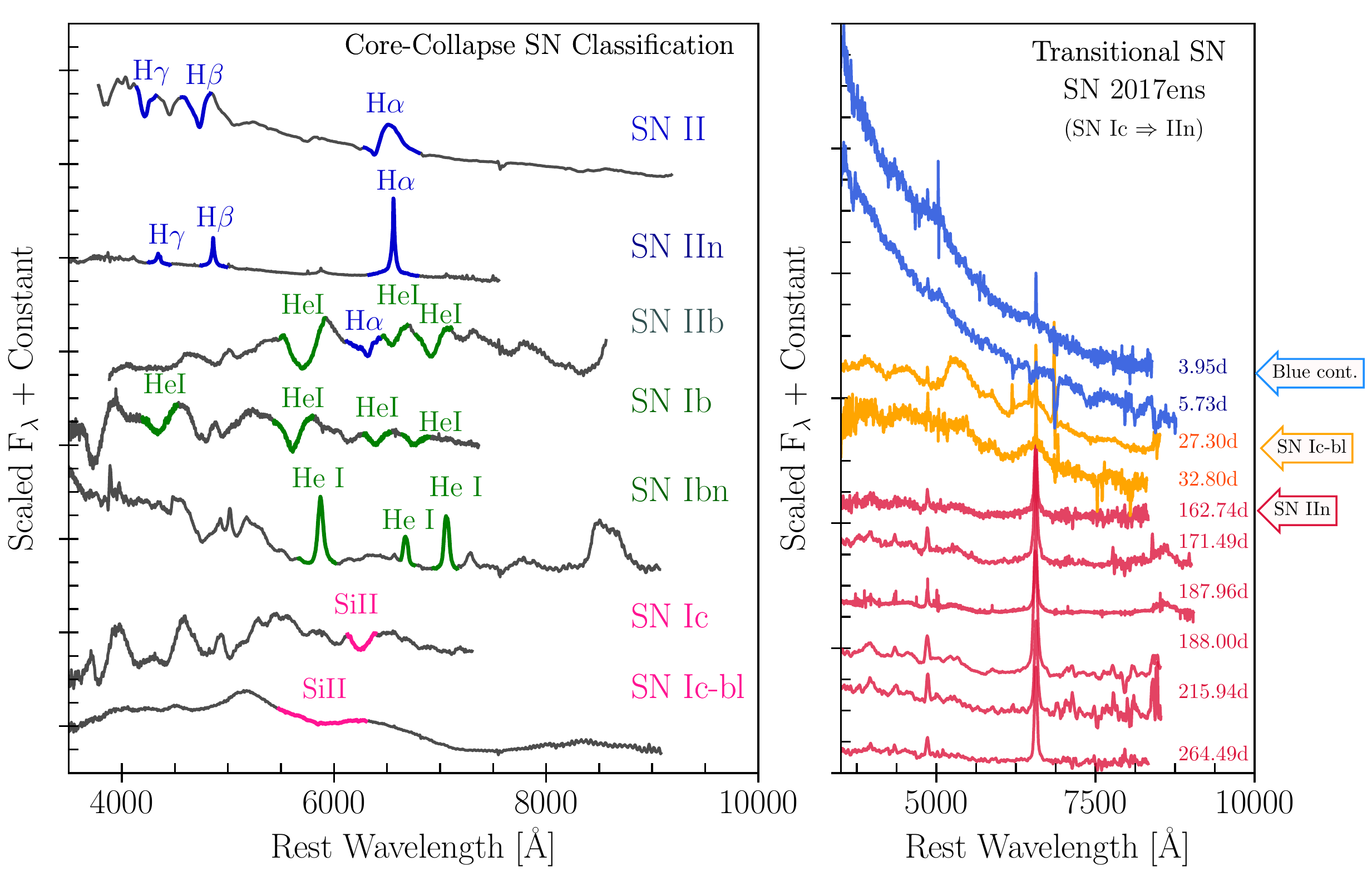} 
\caption{\label{fig:classification} \textbf{Spectral classification of CCSNe.} \textbf{Left:} The main classes of CCSNe are marked based on their composition and line profiles.
The SNe displayed are SN~2007od (SN~II; \citealp{Inserra11, Gutierrez17a}), SN~2010jl (SN~IIn; \citealp{Smith11}), SN~2011hs (SN~IIb; \citealp{Bufano14}), SN~2004gq (SN~Ib; \citealp{Modjaz14}), SN~2006jc (SN~Ibn; \citealp{Pastorello08}), SN~2004gk (SN~Ic; \citealp{Modjaz14}), SN~1998bw (SN~Ic-bl+GRB; \citealp{Patat01}). All spectra are at around maximum light, except that of SN~2006jc, which is at $\sim$4 weeks post-maximum. \textbf{Right:} 
Some objects defy classification as a single type, such as SN~2017ens \citep{Chen18}, shown here, which initially displayed a blue continuum, then the spectrum of a SN~Ic-bl, and then that of a SN~IIn.}. 
\end{figure*}

In Figure~\ref{fig:classification} (left panel), we present spectra around peak brightness of the main CCSN classes. Among them, SNe~II are perhaps the most heterogeneous class with a large range of observed photometric and spectroscopic properties. Because of this diversity, they can be further divided into several subclasses. The first historical subclassification was based on the light curve shape:  SNe showing a linear decline (in magnitudes) in their light curve were named SNe~IIL, while SNe showing a quasi-constant luminosity or a plateau for several weeks were called SNe~IIP \citep{Barbon79}. Later it was discovered that Type IIP and IIL SNe can also be distinguished by a subtle difference in their spectra, with SNe IIP showing deeper absorption in H$\alpha$ compared to SNe IIL \citep{Schlegel96, Gutierrez14, Gutierrez17}.

An additional photometric subclass was later added, namely that of SN~1987A-like SNe. This subclass displays a long (100 $\pm$ 20-day) rise to maximum, following the prototype of SN~1987A \citep[e.g.,][]{Blanco87,Hamuy88} with spectra similar to SNe IIP. 

Based on the spectroscopic properties, further subgroups were later introduced in the SN~II class: SNe~IIb, transitional events between hydrogen-rich SNe~II and hydrogen-poor SNe~Ib (see below), and SNe IIn with relatively narrow emission lines in their spectra produced by dense circumstellar matter (CSM) not yet accelerated by the ejecta \citep{Chevalier81,Fransson82,Schlegel90}. 

\begin{figure*}[ht!]
\includegraphics[width=.97\textwidth]{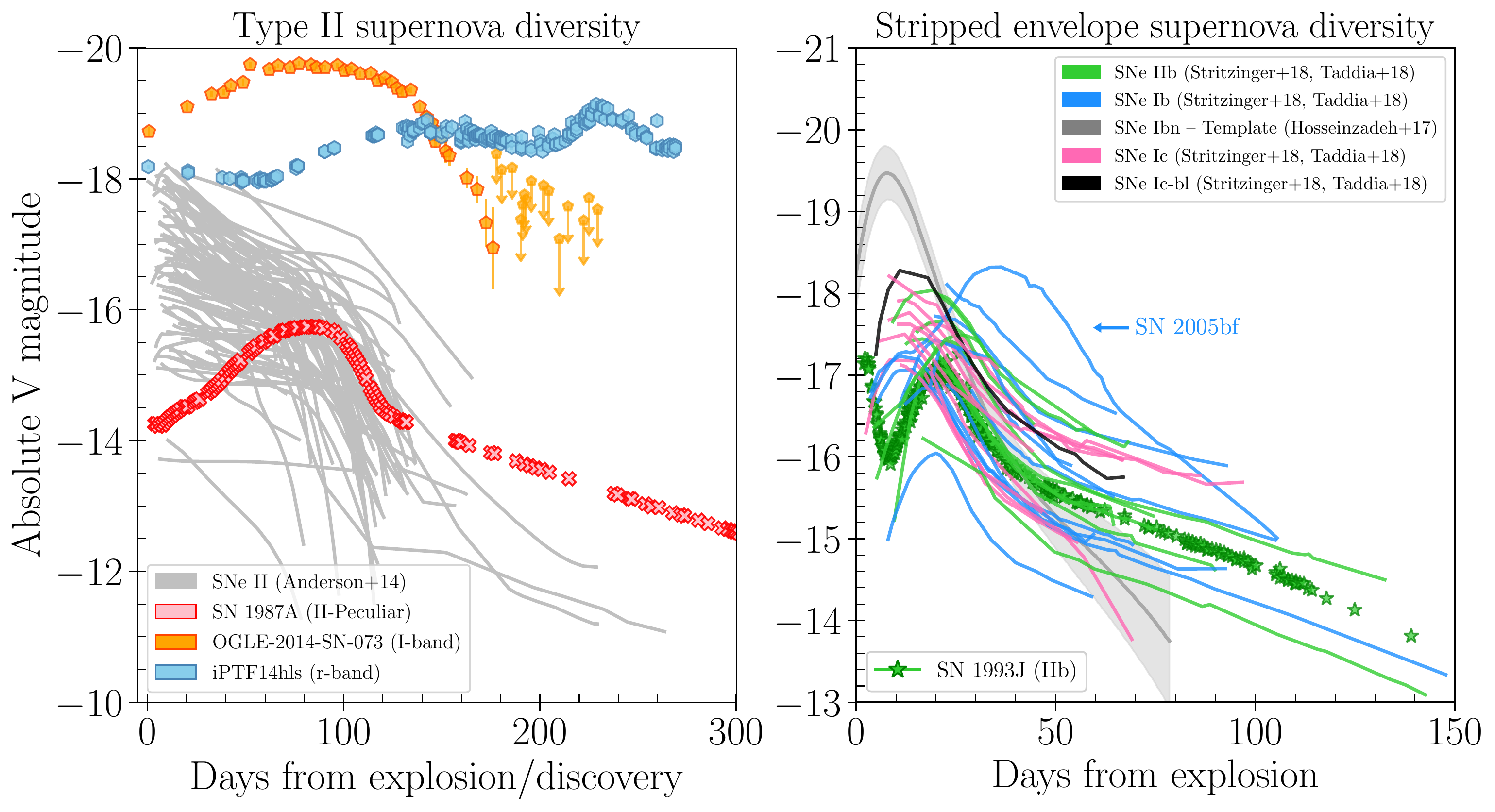} 
\caption{\label{fig:LCs} \textbf{Photometric Diversity of CCSNe.} \textbf{Left:} Example of $V$-band light curves of SNe~II from \citep{Anderson14}, which exhibit a large diversity in luminosity and light curve shape, the drivers of which are not fully understood. Even harder to explain are the power sources of long-lived outliers, such as OGLE-2014-SN-073 \citep{Terreran17} and iPTF14hls (\citealp{Arcavi17}, for which just the first 300 days from discovery are shown here). The light-curve of SN~1987A \citep[e.g., ][]{Blanco87,Hamuy88} is shown in red.
\textbf{Right:} Example of 
$V$-band light curves of stripped envelope SNe, largely from the Carnegie Supernova Project \citep{Stritzinger18, Taddia18_CSPanalysis}, 
which also show a large diversity in luminosity, here thought to result from the different amounts of nickel synthesized in each explosion. The double-peaked light curve of SN~2005bf might be the result of a double-peaked Ni distribution \citep{Tominaga05,Folatelli06} or of a magnetar powering the second peak \citep{Maeda07}. 
We also show the template light curve of SNe Ibn constructed by \citep{Hosseinzadeh17}.
Error bars denote 1$\sigma$ uncertainties. 
The typical uncertainties in the explosion dates for the objects in the right plot are estimated be $\pm$ 3 days \citep{Taddia18_CSPanalysis}. The light curve of the SN IIb SN~1993J is from \citep{Okyudo93,Benson94,Richmond96}}

 \end{figure*}

The subtle spectroscopic division between SNe~IIL and IIP, together with the observation and analysis of intermediate -- between declining (IIL) and plateau (IIP) -- objects (e.g. SN~1992H, \citealp{Clocchiatti96}), questioned the initial separation of SNe~II into IIL and IIP and opened a debate on whether these two subclasses were actually part of a continuum. The first statistical studies of SNe~II with relatively small samples \citep[e.g.,][]{Patat94,Arcavi12,Faran14} found evidence for a separation between IIP and IIL SNe. However, the analysis of larger samples \citep[e.g.,][]{Anderson14, Gutierrez14, Sanders15, Valenti16, Galbany16, Rubin16,  Gutierrez17} shows a continuum in the photometric (and spectroscopic) properties, from plateau to declining light curves (but curiously not from short to long plateaus - i.e. the continuum is in the decline rate of the light curve not in the length of its plateau; see left panel of Figure \ref{fig:LCs}). Specifically ref. \citep{Valenti2015} show that most IIL SNe also have a light curve ``drop'' as seen at the end of the plateau in IIP SN light curves, indicating that cooling and recombination of the hydrogen envelope may power both SNe IIP and IIL light curves for the first few months after explosion, with SNe IIL having an additional power source on top which gradually declines (though the shorter time-scale until the drop in SNe IIL argues for additional intrinsic differences between the sub types).

SN IIP progenitors have been shown (through direct identification in pre-explosion imaging) to be red supergiants \citep[e.g.][]{Smartt15}. There have been no progenitor detections for SNe IIL \citep[except possibly for SN\,2009kr, but its nature is debated;][]{Elias-Rosa10}.
It is therefore not clear which properties differentiate SN IIP and IIL progenitors, but early-time post-SN observations may point at more powerful mass loss winds for SN IIL ones (see below). 
Analysis of larger samples of events with multi-band coverage, together with direct detections of SN IIL progenitors will be key to understanding the IIL/IIP connection.

The Type I SN class is also diverse and divided into a number of subclasses 
(Fig. \ref{fig:classification}):
SNe Ib show strong helium lines, while SNe Ic are the ``rejects'' of the classification scheme, as they are defined by what they do not show: no hydrogen, no helium and no (strong) silicon (distinguishing them from Type Ia SNe which are associated with the explosions of white dwarfs; see the corresponding article on SNe Ia in this issue; \citealp{Jha19_Naturereview}). In the last 20 years, a new subtype has piqued interest: the class of broad-lined SNe Ic (Ic-bl), which is the only subtype connected with long-duration Gamma-Ray Bursts (GRBs), and is among the most powerful explosions in the Universe \citep[see reviews in e.g.,][]{Modjaz11-rev,Cano17_obs_guide}. 
SNe Ic-bl are also related to some of the objects in the emerging field of Superluminous SNe (SLSNe) \citep[e.g.,][]{Liu17,Jerkstrand17}, one of which accompanied an Ultra-long GRB (see the corresponding article on SLSNe in this issue; \citealp{Inserra19_Naturereview}).

Another recent addition to Type I SN diversity is that of SNe Ibn. These events display narrow lines of helium in their spectra and no hydrogen \citep[e.g.,][Fig. \ref{fig:classification}]{Pastorello16,Hosseinzadeh17}, indicating the presence of a hydrogen-poor but helium-rich CSM around the exploding star. Curiously, while SNe IIn (displaying narrow hydrogen lines) show a large diversity of light curve behavior \citep[e.g.,][]{Kiewe2012} including long-lived emission (presumably powered by the interaction of the ejecta with the CSM), SNe Ibn show rather uniform short-lived light curves \citep{Hosseinzadeh17}. Some of these light curves are similar to those of rapidly evolving luminous transients (Fig. \ref{fig:peak}, see \citealp{Inserra19_Naturereview} in this issue).

All of the above classes are associated with CCSNe of massive stars \citep[though recent reports of a SN II and of a SN Ibn in non-star-forming environments challenge this long-held view;][]{Hosseinzadeh2019,Irani2019}, with the SNe IIb, Ib, Ic and Ic-bl subytpes due to different amounts of stripping of the outer hydrogen and helium envelopes of the progenitor thus giving them the collective name ``Stripped-Envelope SNe'' \citep{Clocchiatti97}, or just "Stripped SNe" for short. 

Making matters more complicated, but also interesting, CCSN classification can be time-dependent, with objects changing classes as a function of time ranging from weeks to months to years. One example of an extreme time-dependent classification is that of SN~2017ens \citep[][right panel of Figure~\ref{fig:classification}]{Chen18}, which first showed a featureless blue spectrum (usually seen in Type II SNe shortly after explosion), then transformed to a SN Ic, and then to a SN IIn. This indicates that the SN is illuminating hydrogen-rich material, which the SN progenitor might have expelled before explosion. More time-series spectra are needed for Stripped SNe, especially at later times (months to years) to monitor any type changes due to interaction with a previously expelled envelope. This could help answer a big open question regarding these SNe: how and when are the outer hydrogen and helium layers removed prior to explosion.
 
Like with the IIP and IIL subtypes, another crucial question regarding Stripped SNe is whether SNe IIb, Ib, Ic and Ic-bl are distinct classes or constitute a continuum. Every so often, "transitional objects" are reported. This is not surprising considering that 
massive stars likely do not have either all or none of their outer layers intact. Even the distinction between IIb (some hydrogen layer intact) and Ib (no hydrogen layer) should be more gradual, with objects on a continuum \citep[as indeed indicated by the measurements of e.g.,][]{Liu16}. 

The current classification scheme does not capture the detailed physical state of the pre-explosion star, nor does it allow for a quantitative description of transitional objects. It also does not use all the information in the spectrum, but rather certain features. A new classification scheme which can capture the richness in CCSN diversity is clearly needed. 

Attempts to introduce a new classification scheme that improve upon the old one in some of the ways outlined above have been made over the last few years \citep[e.g.][]{prentice17,sun17} with different degrees of success. The most recent one for Stripped SNe \citep{Williamson19} is the first quantitative one, fulfilling all the needs laid out above, and addressing the aforementioned time-dependent nature of classification. They find that $\sim$2 weeks after peak brightness is the optimal time to differentiate between the different sub classes of Stripped SNe - a result with strong implications for follow-up strategies of current and new-generation SN searches.

\section*{New Peculiar Events and Population Studies of ``Normal'' Events}
~\\

\begin{figure*}[ht!]
\includegraphics[width=.97\textwidth]{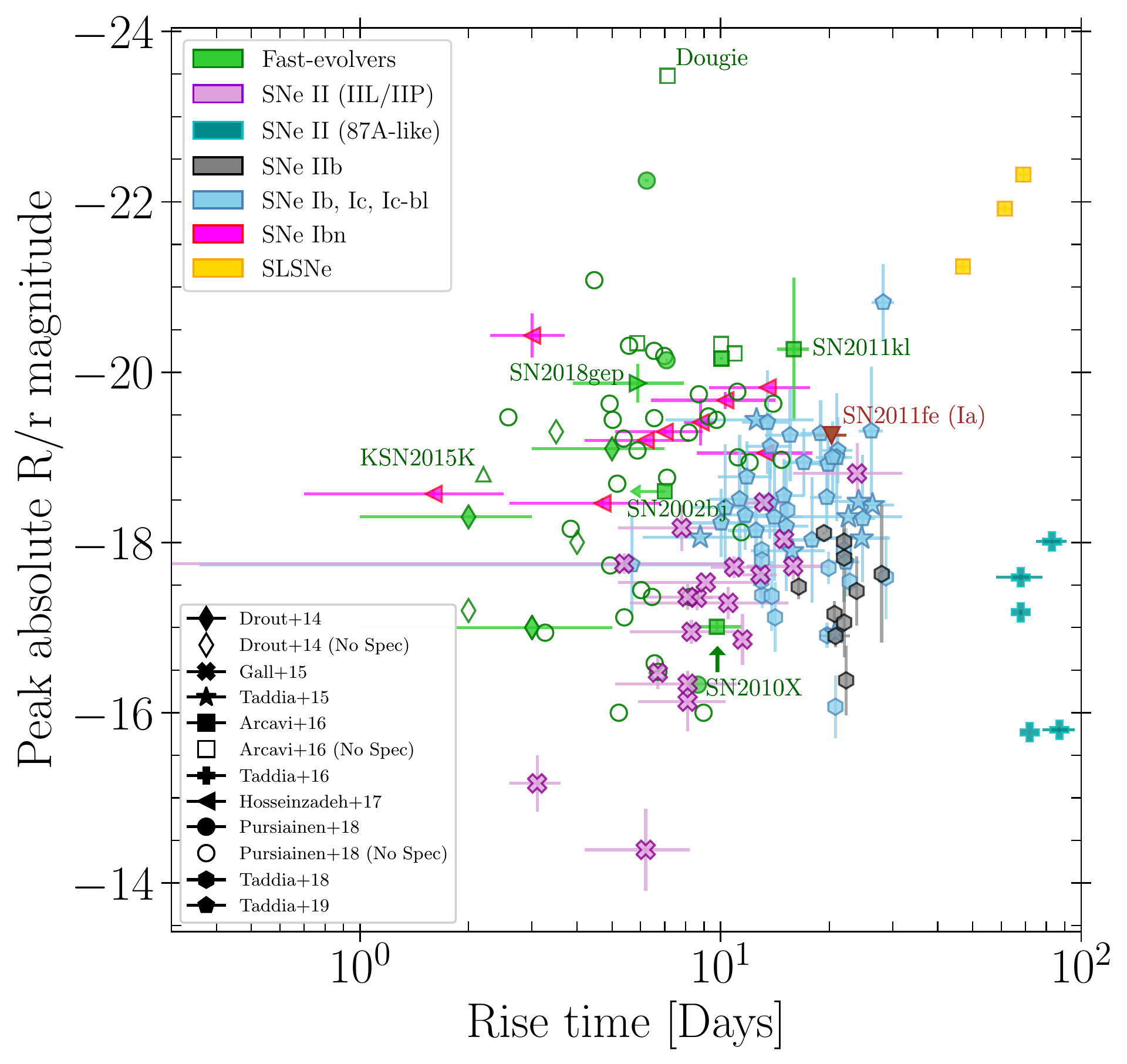} 
\caption{\label{fig:peak} \textbf{A new phase-space: Peak absolute magnitude vs. rise time of CCSNe.}
Such luminosity and time-scale plots can reveal observational biases in the discovery of transients as well as the underlying physics of such events \citep[e.g.][]{Kasliwal2011thesis}. Specifically, the rise time, which can be measured only recently thanks to early discoveries of events with well constrained explosion times, is a proxy for the ejected mass (assuming a central power source and spherical symmetry).
Open symbols denote objects with no spectra, thus their nature as CCSNe is not certain. Shown are: SN~2002bj \citep{Poznanski10}, SN~2010X \citep{Kasliwal11}, Dougie \citep{Vinko15}, SN~2011kl \citep{Greiner15}, KSN2015K \citep{Rest18}, SN~2018gep (\citealp{Ho19}, T. Pritchard et al, in prep.). The markers represent data from different papers/surveys as follows:  \citep{Drout14}: diamond, \citep{Taddia15_sdss}: star, \citep{Gall15}: plus symbol, \citep{Taddia16a}: cross, \citep{Arcavi16}: square, \citep{Taddia19}: pentagon, \citep{Hosseinzadeh17}: left triangle, \citep{Pursiainen18}: circle, \citep{Taddia18_CSPanalysis}: hexagon.
Error bars denote 1$\sigma$ uncertainties. SLSNe stands for Superluminous SNe.}
\end{figure*}

In recent years, wide-field surveys have revealed a large diversity of unusual transients that range from extreme transitional objects (e.g. SN~2017ens; Fig. \ref{fig:classification}; \citealp{Chen18}) to very short-lived rapidly evolving events (see \citealp{Inserra19_Naturereview} in this issue; Fig.~\ref{fig:peak}) to long-lived slowly evolving events, two of which we will briefly mention now. 

iPTF14hls \citep{Arcavi17} and OGLE-2014-SN-073 \citep{Terreran17} were classified as SNe~II based on the presence of hydrogen in their spectra. However, their light curves and spectral evolution are highly unusual. The spectra of iPTF14hls, though identical to those of the IIP subclass, evolved 10 times slower over the course of $\gtrsim600$ days. During the same period, the light curve displayed at least five distinct peaks, a behavior not seen in any other SN. OGLE-2014-SN-073 showed a light curve reminiscent of 1987A-like objects, but $\sim$4 magnitudes brighter than SN~1987A (Fig.~\ref{fig:LCs}), while the spectra showed no evolution during the first $\sim$160 days of follow-up. These two events do not fit into any known class and are hard to explain with current explosion models as canonical Ni-decay power cannot produce the observed properties. Wildly different alternative power sources, such as a magnetar spindown, fallback accretion, electron-positron pair production, hidden interaction with material ejected by the star before its explosion, and jets are being considered \citep[e.g.][]{Arcavi17,Terreran17,Andrews2018,Dessart2018,Soker2018,Wang2018,Woosley2018}. More such events will allow us to better measure their rates and preferred environments and thus perhaps offer more insight into their nature.

Large data rates from wide-field surveys together with large follow-up programs are also enabling population studies of the more common types of events. Figure~\ref{fig:LCs} showcases the diversity of SNe II light curves on the left and the diversity of Stripped SNe light curves on the right. Large-sample population studies are enabling statistical inferences about progenitors and power sources of "normal" run-of-the-mill SNe.

In addition to the samples of SNe II which elucidated the continuum between the IIL and IIP subclasses mentioned above, there have been a number of large data releases of Stripped SN observations over the last few years \citep{Drout11,Bianco14,Modjaz14,Taddia15_sdss,Stritzinger18,Fremling18,Taddia19,Shivvers19,Prentice19}. Works that analyze those large data sets (and in some cases also prior single-object data) report strong constraints on the progenitor systems of Stripped SNe, which is one of the outstanding questions in the field: while using different techniques many works come to the same conclusions, namely that the progenitors of Stripped SNe cannot be only single massive stars but that lower-mass binaries are preferred as the dominant channel \citep[e.g.][]{Smith11,Lyman16,Liu16,Graur17a,Taddia18_CSPanalysis,Prentice19,Taddia19}. However, some nearby objects do not show any trace of a companion star to deep limits, such as Supernova Remnant Cas A in our own Milky Way \citep{Kerzendorf19}, which is known to have been a SN IIb from light echo measurements \citep{Krause08,Rest08}. Thus, the next step for making progress is to determine which individual SNe come from massive single progenitors and which from stars in interacting binaries.

Large samples with early-time data are allowing us to map CCSNe in various phase spaces such as that of rise time and peak luminosity described in Figure \ref{fig:peak}. Since the rise time of a {\bf centrally-powered} SN is roughly related to its ejected mass and the peak luminosity to its power source \cite[e.g.][]{Arnett1982}, such maps can help us find connections between physically similar events, constrain their power sources, and put peculiar events in context. 
 
\section*{Very Early Observations}

~\\

\begin{figure*}[ht!]
\includegraphics[width=.97\textwidth]{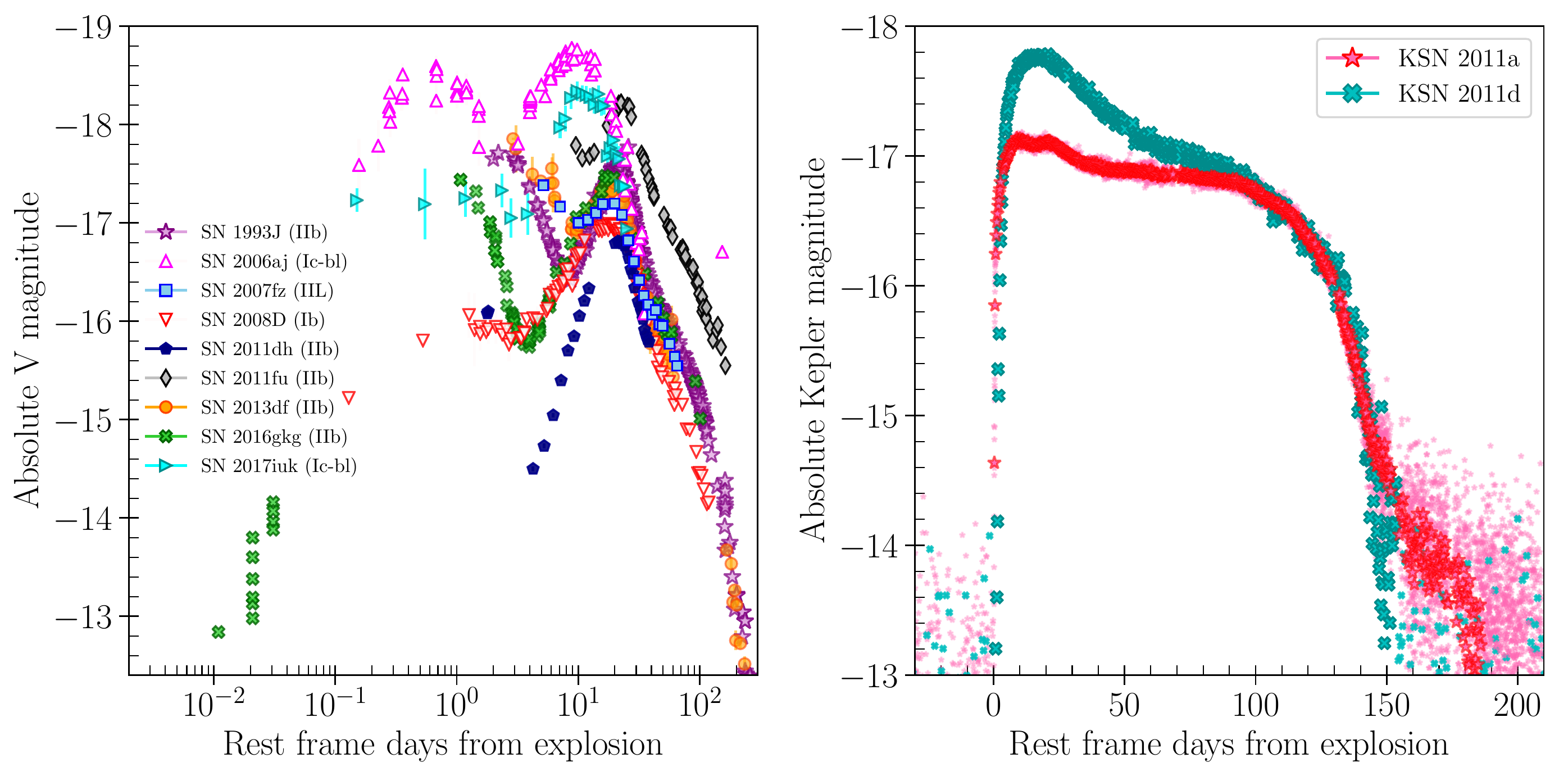} 
\caption{\label{fig:earlyLC}  \textbf{A high-cadence look at CCSN light curves.} \textbf{Left:} Early $V$-band light curves of SNe that show strong evidence for cooling envelope emission as indicated by their double-peaked light curves. Both SNe Ic-bl shown had accompanying GRBs. Objects shown are SN~1993J \citep{Okyudo93,Benson94,Richmond96}, SN~2006aj/GRB060219 \citep{Modjaz06, Bianco14, Brown14}, SN~2007fz \citep{Faran14}, SN~2008D \citep{Mazzali08, Modjaz09, Bianco14}, SN~2011dh \citep{Arcavi11}, SN~2011fu \citep{Kumar13}, SN~2013df \citep{Morales-Garoffolo14, Brown14}, SN~2016gkg \citep{Arcavi17a,Bersten18}, and SN~2017iuk/GRB171205A \citep{Izzo19}. 
 \textbf{Right:} Kepler aperture photometry of two SNe IIP taken with unprecedented cadence and photometric accuracy \citep{Garnavich16}. Data like these can reveal light curve subtleties such as the claim of an optical shock breakout peak at the onset of the SN (not shown here).}
\end{figure*}

An additional new and exciting capability enabled by recent wide-field transient surveys and rapid-response follow-up facilities is that of observing SNe very soon (i.e. hours) after explosion. This allows us to probe the early emission from SNe, which, in stripped events is often powered by mechanisms different than those responsible for the main peak of the light curve (which is powered by the radioactive decay of nickel to cobalt to iron). Two early-time power mechanisms are shock breakout and cooling, as well as interaction with nearby circumstellar material (CSM). The emission produced by these mechanisms is extremely informative. Shock breakout and cooling emission encodes information about the radius and internal density structure of the progenitor right before explosion, and in some cases of the jet propagation physics. Emission from circumstellar interaction, on the other hand, can provide information about the mass loss history of the progenitor in the years or decades before explosion. Together, these insights provide new constraints for modelling the late-stage evolution of massive stars and its connection to their eventual explosion characteristics.

~\\
\noindent{\bf Shock Breakout and Cooling Envelope Emission}
~\\

As the shock generated right after core collapse reaches the edge of the star, where the optical depth $\tau$ is approximately equal to ${c}/{v}$ (with $c$ the speed of light and $v$ the speed of the shock), it will break out, emitting a flash of primarily X-ray or ultraviolet radiation \citep[depending on the radius of the progenitor and its density profile;][]{Colgate1969,Weaver1976,Klein1978,Falk1978,Matzner99,Nakar_Sari_2010,Rabinak_Waxman_2011,Waxman2017}. For large progenitors, the Rayleigh-Jeans tail of the shock breakout emission could extend into the optical \citep[e.g.,][]{Tominaga2011}. The duration of the flash is relatively short ($\lesssim$ hour for a red supergiant).

This time scale is too short for current transient surveys to detect it regularly. Some serendipitous cases of shock breakout detections have been reported in the ultraviolet and in the 
X-ray 
\citep[for the SN~Ib 2008D and the SNe~IIP SNLS-04D2dc and PS1-13arp;][]{Schawinski2008,Soderberg_2008,Gezari2015}, and more recently claimed also to have been seen in the optical see below). The X-ray flash of the SN~Ib 2008D was longer than expected (lasting almost 10 minutes for a stripped-envelope progenitor). The reasons for this are still debated and include asphericity effects \citep{Couch11}, breakout from a thick wind \citep[e.g.,][]{Svirski14}, and the presence of a weak jet \citep{Mazzali08}. The duration of the ultraviolet flash for PS1-13arp was also longer than expected (lasting approximately one day) and is interpreted by \citep{Gezari2015} as shock breakout from a wind.

For GRBs, the nearby and well-observed SN~Ic-bl 2006aj/GRB~060218 showed puzzling X-ray properties, including a thermal component, that have been claimed to be due to shock breakout - either of a shock wave driven by a mildly relativistic shell into the dense wind surrounding the progenitor \citep[e.g.,][]{Campana06} or breakout of a relativistic jet choked by an optically thick envelope \citep[e.g.,][]{Nakar12}. Indeed the latter suggestion has been generalized by \citealp{Nakar12} and collaborators to the whole class of low-luminosity GRBs (LLGRBs) of which SN~2006aj/GRB~060218 is the best example observed. Alternatively, ref. \citep{Irwin16} recently suggested that SN~2006aj/GRB~060218 harbored a long-lived engine that produced a mildly relativistic jet from a non-standard progenitor.

The first claim of an optical detection of shock breakout was for the Type IIP supernova KSN\,2011d \citep{Garnavich_2016} with data from the {\it Kepler} space telescope (originally designed to search for extra-solar planets with exquisite photometric precision and 30-minute cadence; Fig. \ref{fig:earlyLC}). This claim however is challenged by \cite{Rubin2017} who re-analyze the data and declare no statistically significant detection of shock breakout. The second claim is for the Type IIb SN~2016gkg \cite{Bersten18} made thanks to (again, serendipitous) detections by an amateur astronomer (though in this case, the shock breakout peak itself is not observed).

If a massive star is embedded in an optically thick wind at the time of explosion, the shock may continue to propagate through the wind and break out at a radius that is much larger than that of the star. This would increase the time scale of the shock breakout flash and decrease its effective temperature \citep[e.g.,][]{Ofek2010,Chevalier2011,Balberg2011,Svirski2012,Ginzburg2014,Moriya2017} making it easier to observe, especially in the optical. Recently \citep{Forster2018} found signatures of shock breakout in a wind in several CCSNe, indicating that the wind breakout scenario might actually be the more common one and therefore that most massive stars are embedded in a dense wind at the time of explosion. Such a mechanism has also been invoked for some Stripped SNe whose shock breakout time is much longer than the simple photon travel time (e.g. SN Ic-bl 2006aj/GRB 060218 and SN~Ib 2008D - as mentioned above). 

After shock breakout, the ejected material expands and cools, producing radiation known as the ``cooling envelope emission'', which occurs on longer (and thus more easily detectable) time scales. The characteristics of this emission depend most strongly on the radius and internal density structure of the progenitor star just before explosion \citep[e.g.][]{Chevalier08,Nakar_Sari_2010,Rabinak_Waxman_2011,Nakar_Piro_2014,Piro_2015,Sapir_Waxman_2017}. For a red supergiant, for example, the very extended envelope gets ionized by the shock, and slowly recombines across approximately 100 days, producing the plateau in the light curve of SNe IIP. For partially stripped IIb progenitors the cooling envelope emission can last days to approximately a week. 

There are now numerous cases of cooling envelope emission observed in various types of core collapse SNe caught soon after explosion: one SN II \cite{Faran14}, several SNe IIb  \cite{Richmond96,Arcavi11,Kumar13,Bufano14,Morales-Garoffolo14,Arcavi17}, two SNe Ib \citep{Stritzinger02,Mazzali08,Modjaz09}, and two SNe Ic-bl with GRBs  \citep[e.g.][though the latter suggest that the emission is from a cooling jet cocoon, not a cooling stellar envelope]{Campana06,Izzo19}. In all cases the envelope cooling emission is seen as a peak or early excess of the light curves in the bluer filters before the main light curve peak (Fig. \ref{fig:earlyLC}). 

For SNe IIb, the duration, luminosity, and color of the shock cooling peak has been linked to a low mass ($\sim$0.001 -- 0.01 times the mass of the sun, \Msun) extended ($\sim10^{13}\,$cm) envelope surrounding a more compact ($\sim10^{11}$\,cm) core in the progenitor \citep{Hoflich1993,Bersten12,Nakar_Piro_2014,Piro_2015}. Such a structure could be created by binary interaction \citep[e.g.,][]{Benvenuto2013} which might also be responsible for the partial stripping of the hydrogen envelope resulting in the IIb spectral properties of the SNe. This is consistent with other works mentioned above that constrain the dominant channel for Stripped SNe to be binary stellar systems.

~\\
\noindent{\bf Circumstellar Material}
~\\

An additional boost to the emission of SNe can come from interaction between the ejecta and circum-stellar material (CSM) released from the progenitor before explosion. This is most commonly seen in Type IIn SNe where CSM interaction is a dominant power source. 

\begin{figure}[ht!]
\includegraphics[width=.49\textwidth]{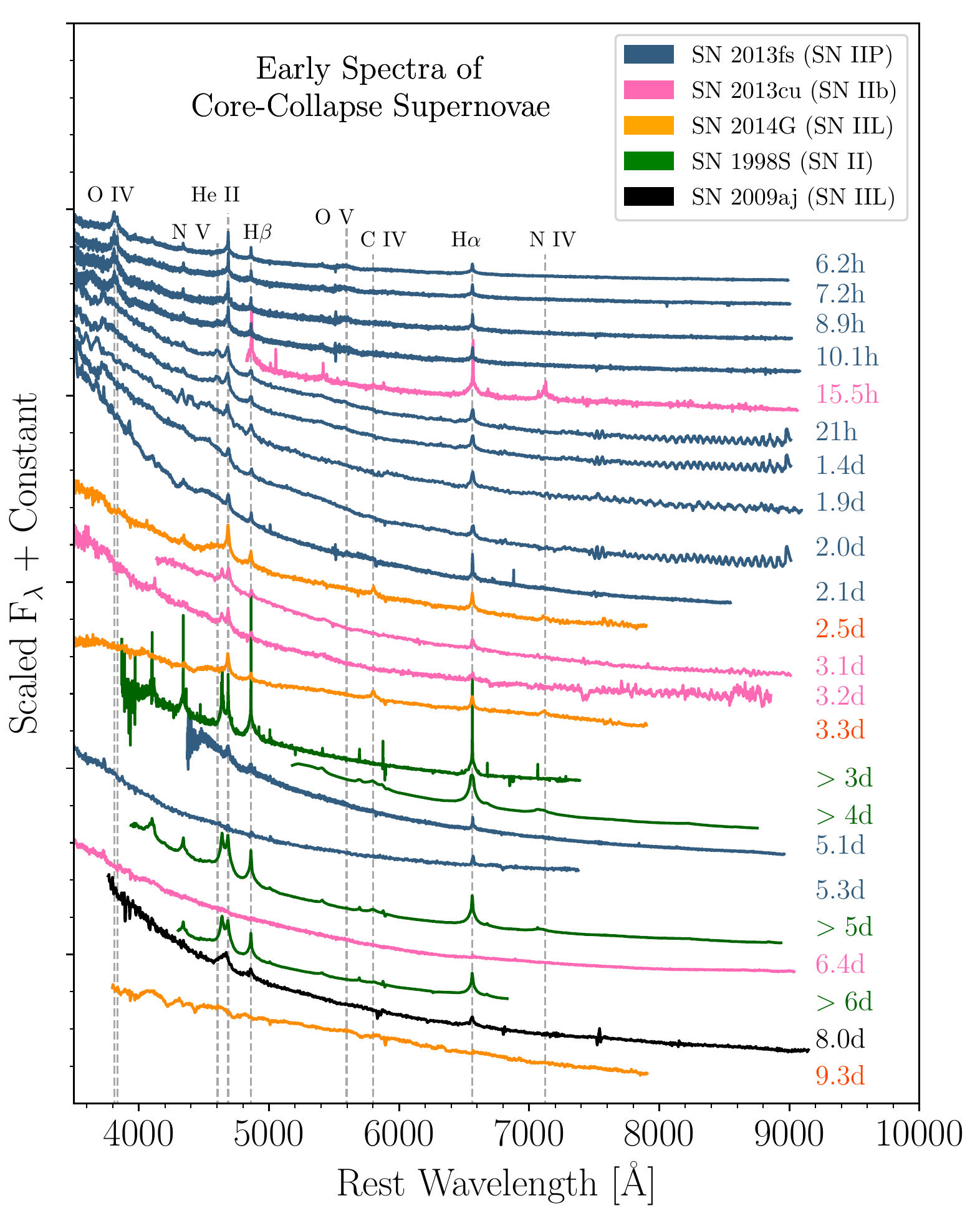} 
\caption{\label{fig:earlyspec} \textbf{``Flash Spectroscopy'' of infant SNe.} Early-time spectra of CCSNe showing narrow lines reveal the presence, composition, and extent of confined CSM ejected by the progenitor shortly before explosion. This information is a crucial constraint on the late stages in stellar evolution models
Shown are spectra of SN~2013cu (pink; \citealp{GalYam14}), SN~1998S (green; \citealp{Shivvers15}), SN~2013fs (blue; \citealp{Yaron17}), SN~2014G (orange; \citealp{Terreran17}), and SN~2009aj (black; \citealp{Gutierrez17a}). Phases with respect to the assumed date of explosion are labeled on the right.}
\end{figure}

Recently though, CSM interaction and/or cooling has been found to be an important source of emission also for other types of SNe, especially at early times. For example, the early light curves of some SNe IIP and IIL cannot be explained with cooling envelope from a bare massive star alone, and require some form of CSM causing the shock breakout to occur at a larger radius \citep[e.g.,][]{Morozova2017,Morozova2018}. 

In fact, a distinguishing factor between SN~IIP and SN~IIL progenitors might be the amount and extent of CSM surrounding the progenitors at the time of explosion, with SN~IIL progenitors exploding in a more massive and extended wind resulting in additional shock cooling emission and brighter early light curves compared with SNe~IIP \citep{Valenti16}.

In addition to early photometry, early spectroscopy can reveal even more about the mass loss history of the progenitor right before explosion. As soon as the shock breaks out of the star, the hot continuum emission could excite the CSM into generating spectral lines. These lines will be relatively narrow since the CSM is released at typical velocities of 10--1000\,km\,s$^{-1}$. There is a time window of a few hours before the most confined CSM, released by the star just months before explosion, is wiped out by the ejecta traveling 10--1000 times faster. In that time window, a spectrum could reveal not just the existence of confined CSM but also its composition and physical extent. 

This method, known as "flash spectroscopy" has been implemented on several SNe II \citep[e.g.][]{GalYam14,Shivvers15,Yaron17}, and has shown that early narrow emission lines are prevalent in most hydrogen-rich SNe \citep{Khazov2016}. This indicates that most massive stars experience some kind of enhanced mass ejection in the months or years prior to core collapse. 

Sequences of early spectra can show the narrow CSM lines slowly broadening as the confined CSM is accelerated by the SN ejecta (Figure~ \ref{fig:earlyspec}), many or which also show enhanced shock cooling emission from their photometry \citep{Morozova2017,Morozova2018}. Sequences of early spectra can show the narrow CSM lines slowly broadening as the confined CSM is accelerated by the SN ejecta (Fig. \ref{fig:earlyspec}). The time scale on which this happens is directly related to the physical extent of the CSM, which is a function of when it was ejected by the star. Therefore these sequences can be used to map the mass loss history of the progenitor in the months or years prior to collapse.

Such spectroscopic data compliments photometric studies of early emission by probing less dense and more extended regions of the CSM. Nevertheless, it again argues that most massive stars experience some kind of enhanced mass ejection in the months or years prior to core collapse.

\section*{The Future: Bright, Fast, and Abundant}
~\\

Despite lots of progress in the quality and quantity of observations of CCSNe, there are still a number of outstanding questions: What are the stellar systems that give rise to these explosions? What are the dominant mechanisms by which the outer layers of Stripped SN progenitors are removed? Which kinds of stars are able to produce SNe with jets and GRBs? How ubiquitous is CSM around CCSN progenitors and what are its properties? 

Fortunately, the number of wide-field transient surveys and follow-up observatories which have enabled progress toward answering these questions continues to grow. The Zwicky Transient Facility \citep[ZTF;][]{ZTF2019} and soon the Large Synoptic Survey Telescope \citep[LSST;][]{LSST2008} are increasing the number of transients discovered by orders of magnitude (ZTF issues $\sim$100,000 transient alerts per night, and LSST is expected to issue $\sim$10 million alerts per night). This promises to increase our sample size and allow us to perform statistical studies on populations of events that until now have been rare. In parallel, small-telescope surveys like the "Distance Less than 40 Mpc" \citep[DLT40;][]{DLT2018} and the "All Sky Automated Survey for SNe" \citep[ASAS-SN;][]{Kochanek2017} are finding young nearby events which can be studied thoroughly on all wavelengths. Seizing the potential of all of these surveys strongly depends on our ability to secure the appropriate follow-up observations.

The advent of such large SN surveys has prompted the development of high throughput low to medium resolution spectrographs like the SED Machine \citep{SEDM2018} and FLOYDS \citep{LCOGT2013} to classify SNe in large numbers. However, more such instruments on larger aperture telescopes are needed for the LSST era. Such a fleet of spectroscopic resources would allow us to increase our samples of events and thus better understand the demographics of "normal" CCSNe as well as uncover more rare (but enlightening) events.

Rapid response followup resources such as the {\it Neil Gehrels Swift Observatory} \citep{Swift2004} and the Las Cumbres Observatory global network of robotic telescopes \citep{LCOGT2013} are paving the way for future ToO-driven facilities. We are now seeing more and more observatories switching to queue observing, and even dynamic scheduling which can adapt on time scales of minutes to accommodate the most urgent and rapidly changing observing constraints. This will allow us to perform flash spectroscopy measurements on samples of events, and thus correlate the composition and mass loss history of massive stars across different parameters such as their metallicity. Dynamic scheduling is also crucial for consistent long-term (months to years) followup of SNe of the kind that enabled the discovery of the "transitional" SN\,2017ens \citep{Chen18}, and the peculiar long-lived SN iPTF14hls \citep{Arcavi17}.

Proposed wide-field space-based ultraviolet imagers \citep{ULTRASAT2014,CUTIE2017}, if launched, will increase the number of discoveries of extremely young SNe and the accuracy with which their progenitor parameters can be determined \citep{Rubin2017}. We will thus not need to rely on luck anymore to see shock breakout and envelope cooling emission, and we will be able to obtain these data regularly for a large number of events. This will allow us to infer the properties of samples of SN progenitors across types and environments.

Finally, we must not forget the importance of repositories storing public data in uniform, easily accessible machine readable formats \citep[such as WISeREP and the Open Supernova Catalog;][]{WISEREP2012,OSC2017} for managing and analyzing these new samples. Systems for coordinating rapid-response observations for multiple targets across various facilities are also crucial \citep[e.g.,][]{TOM2018}. Such systems, together with policies for sharing data publicly are as important as new observing facilities in order for us to be able to reap scientific insights from the continued study of CCSNe in the years to come. 

-----------------

\bibliography{references}

\begin{thebibliography}{100}
\expandafter\ifx\csname url\endcsname\relax
  \def\url#1{\texttt{#1}}\fi
\expandafter\ifx\csname urlprefix\endcsname\relax\def\urlprefix{URL }\fi
\providecommand{\bibinfo}[2]{#2}
\providecommand{\eprint}[2][]{\url{#2}}

\bibitem{SNHandbook}
\bibinfo{author}{{Alsabti}, A.~W.} \& \bibinfo{author}{{Murdin}, P.}
\newblock \emph{\bibinfo{title}{{Handbook of Supernovae}}}
  (\bibinfo{year}{2017}).

\bibitem{Minkowski41}
\bibinfo{author}{{Minkowski}, R.}
\newblock \bibinfo{title}{{Spectra of Supernovae}}.
\newblock \emph{\bibinfo{journal}{\pasp}} \textbf{\bibinfo{volume}{53}},
  \bibinfo{pages}{224} (\bibinfo{year}{1941}).

\bibitem{Filippenko97}
\bibinfo{author}{{Filippenko}, A.~V.}
\newblock \bibinfo{title}{{Optical Spectra of Supernovae}}.
\newblock \emph{\bibinfo{journal}{\araa}} \textbf{\bibinfo{volume}{35}},
  \bibinfo{pages}{309--355} (\bibinfo{year}{1997}).

\bibitem{Gal-Yam16}
\bibinfo{author}{Gal-Yam, A.}
\newblock \bibinfo{title}{Observational and physical classification of
  supernovae}.
\newblock In \emph{\bibinfo{booktitle}{Handbook of Supernovae}},
  \bibinfo{pages}{1--43} (\bibinfo{publisher}{Springer International
  Publishing}, \bibinfo{year}{2016}).
\newblock \urlprefix\url{https://doi.org/10.1007/978-3-319-20794-0_35-1}.

\bibitem{Inserra11}
\bibinfo{author}{{Inserra}, C.} \emph{et~al.}
\newblock \bibinfo{title}{{The Type IIP SN 2007od in UGC 12846: from a bright
  maximum to dust formation in the nebular phase}}.
\newblock \emph{\bibinfo{journal}{\mnras}} \textbf{\bibinfo{volume}{417}},
  \bibinfo{pages}{261--279} (\bibinfo{year}{2011}).

\bibitem{Gutierrez17a}
\bibinfo{author}{{Guti{\'e}rrez}, C.~P.} \emph{et~al.}
\newblock \bibinfo{title}{{Type II Supernova Spectral Diversity. I.
  Observations, Sample Characterization, and Spectral Line Evolution}}.
\newblock \emph{\bibinfo{journal}{\apj}} \textbf{\bibinfo{volume}{850}},
  \bibinfo{pages}{89} (\bibinfo{year}{2017}).

\bibitem{Smith11}
\bibinfo{author}{{Smith}, N.}, \bibinfo{author}{{Li}, W.},
  \bibinfo{author}{{Filippenko}, A.~V.} \& \bibinfo{author}{{Chornock}, R.}
\newblock \bibinfo{title}{{Observed fractions of core-collapse supernova types
  and initial masses of their single and binary progenitor stars}}.
\newblock \emph{\bibinfo{journal}{\mnras}} \textbf{\bibinfo{volume}{412}},
  \bibinfo{pages}{1522--1538} (\bibinfo{year}{2011}).

\bibitem{Bufano14}
\bibinfo{author}{{Bufano}, F.} \emph{et~al.}
\newblock \bibinfo{title}{{SN 2011hs: a fast and faint Type IIb supernova from
  a supergiant progenitor}}.
\newblock \emph{\bibinfo{journal}{\mnras}} \textbf{\bibinfo{volume}{439}},
  \bibinfo{pages}{1807--1828} (\bibinfo{year}{2014}).

\bibitem{Modjaz14}
\bibinfo{author}{{Modjaz}, M.} \emph{et~al.}
\newblock \bibinfo{title}{{Optical Spectra of 73 Stripped-envelope
  Core-collapse Supernovae}}.
\newblock \emph{\bibinfo{journal}{\aj}} \textbf{\bibinfo{volume}{147}},
  \bibinfo{pages}{99} (\bibinfo{year}{2014}).

\bibitem{Pastorello08}
\bibinfo{author}{{Pastorello}, A.} \emph{et~al.}
\newblock \bibinfo{title}{{Massive stars exploding in a He-rich circumstellar
  medium - I. Type Ibn (SN 2006jc-like) events}}.
\newblock \emph{\bibinfo{journal}{\mnras}} \textbf{\bibinfo{volume}{389}},
  \bibinfo{pages}{113--130} (\bibinfo{year}{2008}).

\bibitem{Patat01}
\bibinfo{author}{{Patat}, F.} \emph{et~al.}
\newblock \bibinfo{title}{{The Metamorphosis of SN 1998bw}}.
\newblock \emph{\bibinfo{journal}{\apj}} \textbf{\bibinfo{volume}{555}},
  \bibinfo{pages}{900--917} (\bibinfo{year}{2001}).

\bibitem{Chen18}
\bibinfo{author}{{Chen}, T.~W.} \emph{et~al.}
\newblock \bibinfo{title}{{SN 2017ens: The Metamorphosis of a Luminous
  Broadlined Type Ic Supernova into an SN IIn}}.
\newblock \emph{\bibinfo{journal}{\apj}} \textbf{\bibinfo{volume}{867}},
  \bibinfo{pages}{L31} (\bibinfo{year}{2018}).

\bibitem{Barbon79}
\bibinfo{author}{{Barbon}, R.}, \bibinfo{author}{{Ciatti}, F.} \&
  \bibinfo{author}{{Rosino}, L.}
\newblock \bibinfo{title}{{Photometric properties of type II supernovae}}.
\newblock \emph{\bibinfo{journal}{\aap}} \textbf{\bibinfo{volume}{72}},
  \bibinfo{pages}{287--292} (\bibinfo{year}{1979}).

\bibitem{Schlegel96}
\bibinfo{author}{{Schlegel}, E.~M.}
\newblock \bibinfo{title}{{On the Early Spectroscopic Distinction of Type II
  Supernovae}}.
\newblock \emph{\bibinfo{journal}{\aj}} \textbf{\bibinfo{volume}{111}},
  \bibinfo{pages}{1660} (\bibinfo{year}{1996}).

\bibitem{Gutierrez14}
\bibinfo{author}{{Guti{\'e}rrez}, C.~P.} \emph{et~al.}
\newblock \bibinfo{title}{{Halpha Spectral Diversity of Type II Supernovae: Correlations with Photometric Properties}}.
\newblock \emph{\bibinfo{journal}{\apjl}} \textbf{\bibinfo{volume}{786}},
  \bibinfo{pages}{L15} (\bibinfo{year}{2014}).

\bibitem{Gutierrez17}
\bibinfo{author}{{Guti{\'e}rrez}, C.~P.} \emph{et~al.}
\newblock \bibinfo{title}{{Type II Supernova Spectral Diversity. II.
  Spectroscopic and Photometric Correlations}}.
\newblock \emph{\bibinfo{journal}{\apj}} \textbf{\bibinfo{volume}{850}},
  \bibinfo{pages}{90} (\bibinfo{year}{2017}).

\bibitem{Blanco87}
\bibinfo{author}{{Blanco}, V.~M.} \emph{et~al.}
\newblock \bibinfo{title}{{Supernova 1987A in the Large Magellanic Cloud -
  Initial observations at Cerro Tololo}}.
\newblock \emph{\bibinfo{journal}{\apj}} \textbf{\bibinfo{volume}{320}},
  \bibinfo{pages}{589--596} (\bibinfo{year}{1987}).

\bibitem{Hamuy88}
\bibinfo{author}{{Hamuy}, M.}, \bibinfo{author}{{Suntzeff}, N.~B.},
  \bibinfo{author}{{Gonzalez}, R.} \& \bibinfo{author}{{Martin}, G.}
\newblock \bibinfo{title}{{SN 1987A in the LMC - UBVRI photometry at Cerro
  Tololo}}.
\newblock \emph{\bibinfo{journal}{\aj}} \textbf{\bibinfo{volume}{95}},
  \bibinfo{pages}{63--83} (\bibinfo{year}{1988}).

\bibitem{Chevalier81}
\bibinfo{author}{{Chevalier}, R.~A.}
\newblock \bibinfo{title}{{The interaction of the radiation from a type II
  supernova with a circumstellar shell.}}
\newblock \emph{\bibinfo{journal}{\apj}} \textbf{\bibinfo{volume}{251}},
  \bibinfo{pages}{259--265} (\bibinfo{year}{1981}).

\bibitem{Fransson82}
\bibinfo{author}{{Fransson}, C.}
\newblock \bibinfo{title}{{X-ray and UV-emission from supernova shock waves in
  stellar winds}}.
\newblock \emph{\bibinfo{journal}{\aap}} \textbf{\bibinfo{volume}{111}},
  \bibinfo{pages}{140--150} (\bibinfo{year}{1982}).

\bibitem{Schlegel90}
\bibinfo{author}{{Schlegel}, E.~M.}
\newblock \bibinfo{title}{{A new subclass of Type II supernovae?}}
\newblock \emph{\bibinfo{journal}{\mnras}} \textbf{\bibinfo{volume}{244}},
  \bibinfo{pages}{269--271} (\bibinfo{year}{1990}).

\bibitem{Clocchiatti96}
\bibinfo{author}{{Clocchiatti}, A.} \emph{et~al.}
\newblock \bibinfo{title}{{A Study of SN 1992H in NGC 5377}}.
\newblock \emph{\bibinfo{journal}{\aj}} \textbf{\bibinfo{volume}{111}},
  \bibinfo{pages}{1286} (\bibinfo{year}{1996}).

\bibitem{Patat94}
\bibinfo{author}{{Patat}, F.}, \bibinfo{author}{{Barbon}, R.},
  \bibinfo{author}{{Cappellaro}, E.} \& \bibinfo{author}{{Turatto}, M.}
\newblock \bibinfo{title}{{Light curves of type II supernovae. 2: The
  analysis}}.
\newblock \emph{\bibinfo{journal}{\aap}} \textbf{\bibinfo{volume}{282}},
  \bibinfo{pages}{731--741} (\bibinfo{year}{1994}).

\bibitem{Arcavi12}
\bibinfo{author}{{Arcavi}, I.} \emph{et~al.}
\newblock \bibinfo{title}{{Caltech Core-Collapse Project (CCCP) Observations of
  Type II Supernovae: Evidence for Three Distinct Photometric Subtypes}}.
\newblock \emph{\bibinfo{journal}{\apjl}} \textbf{\bibinfo{volume}{756}},
  \bibinfo{pages}{L30} (\bibinfo{year}{2012}).

\bibitem{Faran14}
\bibinfo{author}{{Faran}, T.} \emph{et~al.}
\newblock \bibinfo{title}{{A sample of Type II-L supernovae}}.
\newblock \emph{\bibinfo{journal}{\mnras}} \textbf{\bibinfo{volume}{445}},
  \bibinfo{pages}{554--569} (\bibinfo{year}{2014}).

\bibitem{Anderson14}
\bibinfo{author}{{Anderson}, J.~P.} \emph{et~al.}
\newblock \bibinfo{title}{{Characterizing the V-band Light-curves of
  Hydrogen-rich Type II Supernovae}}.
\newblock \emph{\bibinfo{journal}{\apj}} \textbf{\bibinfo{volume}{786}},
  \bibinfo{pages}{67} (\bibinfo{year}{2014}).

\bibitem{Sanders15}
\bibinfo{author}{{Sanders}, N.~E.} \emph{et~al.}
\newblock \bibinfo{title}{{Toward Characterization of the Type IIP Supernova
  Progenitor Population: A Statistical Sample of Light Curves from
  Pan-STARRS1}}.
\newblock \emph{\bibinfo{journal}{\apj}} \textbf{\bibinfo{volume}{799}},
  \bibinfo{pages}{208} (\bibinfo{year}{2015}).

\bibitem{Valenti16}
\bibinfo{author}{{Valenti}, S.} \emph{et~al.}
\newblock \bibinfo{title}{{The diversity of Type II supernova versus the
  similarity in their progenitors}}.
\newblock \emph{\bibinfo{journal}{\mnras}} \textbf{\bibinfo{volume}{459}},
  \bibinfo{pages}{3939--3962} (\bibinfo{year}{2016}).

\bibitem{Galbany16}
\bibinfo{author}{{Galbany}, L.} \emph{et~al.}
\newblock \bibinfo{title}{{UBVRIz Light Curves of 51 Type II Supernovae}}.
\newblock \emph{\bibinfo{journal}{\aj}} \textbf{\bibinfo{volume}{151}},
  \bibinfo{pages}{33} (\bibinfo{year}{2016}).

\bibitem{Rubin16}
\bibinfo{author}{{Rubin}, A.} \emph{et~al.}
\newblock \bibinfo{title}{{Type II Supernova Energetics and Comparison of Light
  Curves to Shock-cooling Models}}.
\newblock \emph{\bibinfo{journal}{\apj}} \textbf{\bibinfo{volume}{820}},
  \bibinfo{pages}{33} (\bibinfo{year}{2016}).

\bibitem{Valenti2015}
\bibinfo{author}{{Valenti}, S.} \emph{et~al.}
\newblock \bibinfo{title}{{Supernova 2013by: a Type IIL supernova with a
  IIP-like light-curve drop}}.
\newblock \emph{\bibinfo{journal}{\mnras}} \textbf{\bibinfo{volume}{448}},
  \bibinfo{pages}{2608--2616} (\bibinfo{year}{2015}).

\bibitem{Smartt15}
\bibinfo{author}{{Smartt}, S.~J.}
\newblock \bibinfo{title}{{Observational Constraints on the Progenitors of
  Core-Collapse Supernovae: The Case for Missing High-Mass Stars}}.
\newblock \emph{\bibinfo{journal}{\pasa}} \textbf{\bibinfo{volume}{32}},
  \bibinfo{pages}{e016} (\bibinfo{year}{2015}).

\bibitem{Elias-Rosa10}
\bibinfo{author}{{Elias-Rosa}, N.} \emph{et~al.}
\newblock \bibinfo{title}{{The Massive Progenitor of the Type II-linear
  Supernova 2009kr}}.
\newblock \emph{\bibinfo{journal}{\apjl}} \textbf{\bibinfo{volume}{714}},
  \bibinfo{pages}{L254--L259} (\bibinfo{year}{2010}).

\bibitem{Terreran17}
\bibinfo{author}{{Terreran}, G.} \emph{et~al.}
\newblock \bibinfo{title}{{Hydrogen-rich supernovae beyond the neutrino-driven
  core-collapse paradigm}}.
\newblock \emph{\bibinfo{journal}{Nature Astronomy}}
  \textbf{\bibinfo{volume}{1}}, \bibinfo{pages}{713--720}
  (\bibinfo{year}{2017}).

\bibitem{Arcavi17}
\bibinfo{author}{{Arcavi}, I.} \emph{et~al.}
\newblock \bibinfo{title}{{Energetic eruptions leading to a peculiar
  hydrogen-rich explosion of a massive star}}.
\newblock \emph{\bibinfo{journal}{\nat}} \textbf{\bibinfo{volume}{551}},
  \bibinfo{pages}{210--213} (\bibinfo{year}{2017}).

\bibitem{Stritzinger18}
\bibinfo{author}{{Stritzinger}, M.~D.} \emph{et~al.}
\newblock \bibinfo{title}{{The Carnegie Supernova Project I. Photometry data
  release of low-redshift stripped-envelope supernovae}}.
\newblock \emph{\bibinfo{journal}{\aap}} \textbf{\bibinfo{volume}{609}},
  \bibinfo{pages}{A134} (\bibinfo{year}{2018}).

\bibitem{Taddia18_CSPanalysis}
\bibinfo{author}{{Taddia}, F.} \emph{et~al.}
\newblock \bibinfo{title}{{The Carnegie Supernova Project I. Analysis of
  stripped-envelope supernova light curves}}.
\newblock \emph{\bibinfo{journal}{\aap}} \textbf{\bibinfo{volume}{609}},
  \bibinfo{pages}{A136} (\bibinfo{year}{2018}).

\bibitem{Tominaga05}
\bibinfo{author}{{Tominaga}, N.} \emph{et~al.}
\newblock \bibinfo{title}{{The Unique Type Ib Supernova 2005bf: A WN Star
  Explosion Model for Peculiar Light Curves and Spectra}}.
\newblock \emph{\bibinfo{journal}{\apj}} \textbf{\bibinfo{volume}{633}},
  \bibinfo{pages}{L97--L100} (\bibinfo{year}{2005}).

\bibitem{Folatelli06}
\bibinfo{author}{{Folatelli}, G.} \emph{et~al.}
\newblock \bibinfo{title}{{SN 2005bf: A Possible Transition Event between Type
  Ib/c Supernovae and Gamma-Ray Bursts}}.
\newblock \emph{\bibinfo{journal}{\apj}} \textbf{\bibinfo{volume}{641}},
  \bibinfo{pages}{1039--1050} (\bibinfo{year}{2006}).

\bibitem{Maeda07}
\bibinfo{author}{{Maeda}, K.} \emph{et~al.}
\newblock \bibinfo{title}{{The Unique Type Ib Supernova 2005bf at Nebular
  Phases: A Possible Birth Event of a Strongly Magnetized Neutron Star}}.
\newblock \emph{\bibinfo{journal}{\apj}} \textbf{\bibinfo{volume}{666}},
  \bibinfo{pages}{1069--1082} (\bibinfo{year}{2007}).

\bibitem{Hosseinzadeh17}
\bibinfo{author}{{Hosseinzadeh}, G.} \emph{et~al.}
\newblock \bibinfo{title}{{Type Ibn Supernovae Show Photometric Ho mogeneity
  and Spectral Diversity at Maximum Light}}.
\newblock \emph{\bibinfo{journal}{\apj}} \textbf{\bibinfo{volume}{836}},
  \bibinfo{pages}{158} (\bibinfo{year}{2017}).

\bibitem{Okyudo93}
\bibinfo{author}{{Okyudo}, M.}, \bibinfo{author}{{Kato}, T.},
  \bibinfo{author}{{Ishida}, T.}, \bibinfo{author}{{Tokimasa}, N.} \&
  \bibinfo{author}{{Yamaoka}, H.}
\newblock \bibinfo{title}{{A V-Band Light Curve of SN 1993J during the First 50
  Days}}.
\newblock \emph{\bibinfo{journal}{Publications of the Astronomical Society of
  Japan}} \textbf{\bibinfo{volume}{45}}, \bibinfo{pages}{L63--L65}
  (\bibinfo{year}{1993}).

\bibitem{Benson94}
\bibinfo{author}{{Benson}, P.~J.} \emph{et~al.}
\newblock \bibinfo{title}{{Light Curves of SN 1993J From The Keck Northeast
  Astronomy Consortium}}.
\newblock \emph{\bibinfo{journal}{\aj}} \textbf{\bibinfo{volume}{107}},
  \bibinfo{pages}{1453} (\bibinfo{year}{1994}).

\bibitem{Richmond96}
\bibinfo{author}{{Richmond}, M.~W.}, \bibinfo{author}{{Treffers}, R.~R.},
  \bibinfo{author}{{Filippenko}, A.~V.} \& \bibinfo{author}{{Paik}, Y.}
\newblock \bibinfo{title}{{UBVRI Photometry of SN 1993J in M81: Days 3 to
  365}}.
\newblock \emph{\bibinfo{journal}{\aj}} \textbf{\bibinfo{volume}{112}},
  \bibinfo{pages}{732} (\bibinfo{year}{1996}).

\bibitem{Jha19_Naturereview}
\bibinfo{author}{{Jha}, A.}, \bibinfo{author}{{Maguire}, K.} \&
  \bibinfo{author}{{Sullivan}, M.}
\newblock \bibinfo{title}{{}}.
\newblock \emph{\bibinfo{journal}{\natastro}}  (\bibinfo{year}{2019}).

\bibitem{Modjaz11-rev}
\bibinfo{author}{{Modjaz}, M.}
\newblock \bibinfo{title}{{Stellar forensics with the Supernova-GRB
  connection}}.
\newblock \emph{\bibinfo{journal}{Astronomische Nachrichten}}
  \textbf{\bibinfo{volume}{332}}, \bibinfo{pages}{434--447}
  (\bibinfo{year}{2011}).

\bibitem{Cano17_obs_guide}
\bibinfo{author}{Cano, Z.}, \bibinfo{author}{Wang, S.-Q.},
  \bibinfo{author}{Dai, Z.-G.} \& \bibinfo{author}{Wu, X.-F.}
\newblock \bibinfo{title}{The observer's guide to the gamma-ray burst supernova
  connection}.
\newblock \emph{\bibinfo{journal}{Advances in Astronomy}}
  \textbf{\bibinfo{volume}{2017}}, \bibinfo{pages}{1--41}
  (\bibinfo{year}{2017}).
\newblock \urlprefix\url{https://doi.org/10.1155/2017/8929054}.

\bibitem{Liu17}
\bibinfo{author}{{Liu}, Y.-Q.}, \bibinfo{author}{{Modjaz}, M.} \&
  \bibinfo{author}{{Bianco}, F.~B.}
\newblock \bibinfo{title}{{Analyzing the Largest Spectroscopic Data Set of
  Hydrogen-poor Super-luminous Supernovae}}.
\newblock \emph{\bibinfo{journal}{\apj}} \textbf{\bibinfo{volume}{845}},
  \bibinfo{pages}{85} (\bibinfo{year}{2017}).

\bibitem{Jerkstrand17}
\bibinfo{author}{{Jerkstrand}, A.} \emph{et~al.}
\newblock \bibinfo{title}{{Long-duration Superluminous Supernovae at Late
  Times}}.
\newblock \emph{\bibinfo{journal}{\apj}} \textbf{\bibinfo{volume}{835}},
  \bibinfo{pages}{13} (\bibinfo{year}{2017}).

\bibitem{Inserra19_Naturereview}
\bibinfo{author}{{Inserra}, C.}
\newblock \bibinfo{title}{{}}.
\newblock \emph{\bibinfo{journal}{\natastro}}  (\bibinfo{year}{2019}).

\bibitem{Pastorello16}
\bibinfo{author}{{Pastorello}, A.} \emph{et~al.}
\newblock \bibinfo{title}{{Massive stars exploding in a He-rich circumstellar
  medium - IX. SN 2014av, and characterization of Type Ibn SNe}}.
\newblock \emph{\bibinfo{journal}{\mnras}} \textbf{\bibinfo{volume}{456}},
  \bibinfo{pages}{853--869} (\bibinfo{year}{2016}).

\bibitem{Kiewe2012}
\bibinfo{author}{{Kiewe}, M.} \emph{et~al.}
\newblock \bibinfo{title}{{Caltech Core-Collapse Project (CCCP) Observations of
  Type IIn Supernovae: Typical Properties and Implications for Their Progenitor
  Stars}}.
\newblock \emph{\bibinfo{journal}{\apj}} \textbf{\bibinfo{volume}{744}},
  \bibinfo{pages}{10} (\bibinfo{year}{2012}).

\bibitem{Hosseinzadeh2019}
\bibinfo{author}{{Hosseinzadeh}, G.} \emph{et~al.}
\newblock \bibinfo{title}{{Type Ibn Supernovae May not all Come from Massive
  Stars}}.
\newblock \emph{\bibinfo{journal}{\apjl}} \textbf{\bibinfo{volume}{871}},
  \bibinfo{pages}{L9} (\bibinfo{year}{2019}).

\bibitem{Irani2019}
\bibinfo{author}{{Irani}, I.} \emph{et~al.}
\newblock \bibinfo{title}{{SN 2016hil-- a Type II supernova in the remote
  outskirts of an elliptical host and its origin}}.
\newblock \emph{\bibinfo{journal}{submitted to \apj (arXiv:1904.01425)}}
  (\bibinfo{year}{2019}).

\bibitem{Clocchiatti97}
\bibinfo{author}{{Clocchiatti}, A.} \emph{et~al.}
\newblock \bibinfo{title}{{SN 1983V in NGC 1365 and the Nature of Stripped
  Envelope Core-Collapse Supernovae}}.
\newblock \emph{\bibinfo{journal}{\apj}} \textbf{\bibinfo{volume}{483}},
  \bibinfo{pages}{675--+} (\bibinfo{year}{1997}).

\bibitem{Liu16}
\bibinfo{author}{{Liu}, Y.-Q.}, \bibinfo{author}{{Modjaz}, M.},
  \bibinfo{author}{{Bianco}, F.~B.} \& \bibinfo{author}{{Graur}, O.}
\newblock \bibinfo{title}{{Analyzing the Largest Spectroscopic Data Set of
  Stripped Supernovae to Improve Their Identifications and Constrain Their
  Progenitors}}.
\newblock \emph{\bibinfo{journal}{\apj}} \textbf{\bibinfo{volume}{827}},
  \bibinfo{pages}{90} (\bibinfo{year}{2016}).

\bibitem{prentice17}
\bibinfo{author}{{Prentice}, S.~J.} \& \bibinfo{author}{{Mazzali}, P.~A.}
\newblock \bibinfo{title}{{A physically motivated classification of
  stripped-envelope supernovae}}.
\newblock \emph{\bibinfo{journal}{\mnras}} \textbf{\bibinfo{volume}{469}},
  \bibinfo{pages}{2672--2694} (\bibinfo{year}{2017}).

\bibitem{sun17}
\bibinfo{author}{{Sun}, F.} \& \bibinfo{author}{{Gal-Yam}, A.}
\newblock \bibinfo{title}{{Quantitative Classification of Type I Supernovae
  Using Spectroscopic Features at Maximum Brightness}}.
\newblock \emph{\bibinfo{journal}{arXiv e-prints}}  (\bibinfo{year}{2017}).

\bibitem{Williamson19}
\bibinfo{author}{{Williamson}, M.}, \bibinfo{author}{{Modjaz}, M.} \&
  \bibinfo{author}{{Bianco}, F.}
\newblock \bibinfo{title}{{Optimal Classification and Outlier Detection for
  Stripped-Envelope Core-Collapse Supernovae}}.
\newblock \emph{\bibinfo{journal}{ApJ, submitted (arXiv:1903.06815)}}
  (\bibinfo{year}{2019}).

\bibitem{Andrews2018}
\bibinfo{author}{{Andrews}, J.~E.} \& \bibinfo{author}{{Smith}, N.}
\newblock \bibinfo{title}{{Strong late-time circumstellar interaction in the
  peculiar supernova iPTF14hls}}.
\newblock \emph{\bibinfo{journal}{\mnras}} \textbf{\bibinfo{volume}{477}},
  \bibinfo{pages}{74--79} (\bibinfo{year}{2018}).

\bibitem{Dessart2018}
\bibinfo{author}{{Dessart}, L.}
\newblock \bibinfo{title}{{A magnetar model for the hydrogen-rich
  super-luminous supernova iPTF14hls}}.
\newblock \emph{\bibinfo{journal}{\aap}} \textbf{\bibinfo{volume}{610}},
  \bibinfo{pages}{L10} (\bibinfo{year}{2018}).

\bibitem{Soker2018}
\bibinfo{author}{{Soker}, N.} \& \bibinfo{author}{{Gilkis}, A.}
\newblock \bibinfo{title}{{Explaining iPTF14hls as a common-envelope jets
  supernova}}.
\newblock \emph{\bibinfo{journal}{\mnras}} \textbf{\bibinfo{volume}{475}},
  \bibinfo{pages}{1198--1202} (\bibinfo{year}{2018}).

\bibitem{Wang2018}
\bibinfo{author}{{Wang}, L.~J.} \emph{et~al.}
\newblock \bibinfo{title}{{A Fallback Accretion Model for the Unusual Type II-P
  Supernova iPTF14hls}}.
\newblock \emph{\bibinfo{journal}{\apj}} \textbf{\bibinfo{volume}{865}},
  \bibinfo{pages}{95} (\bibinfo{year}{2018}).

\bibitem{Woosley2018}
\bibinfo{author}{{Woosley}, S.~E.}
\newblock \bibinfo{title}{{Models for the Unusual Supernova iPTF14hls}}.
\newblock \emph{\bibinfo{journal}{\apj}} \textbf{\bibinfo{volume}{863}},
  \bibinfo{pages}{105} (\bibinfo{year}{2018}).

\bibitem{Kasliwal2011thesis}
\bibinfo{author}{{Kasliwal}, M.~M.}
\newblock \emph{\bibinfo{title}{{Bridging the gap : elusive explosions in the
  local universe}}}.
\newblock Ph.D. thesis, \bibinfo{school}{California Institute of Technology}
  (\bibinfo{year}{2011}).

\bibitem{Poznanski10}
\bibinfo{author}{{Poznanski}, D.} \emph{et~al.}
\newblock \bibinfo{title}{{An Unusually Fast-Evolving Supernova}}.
\newblock \emph{\bibinfo{journal}{Science}} \textbf{\bibinfo{volume}{327}},
  \bibinfo{pages}{58} (\bibinfo{year}{2010}).

\bibitem{Kasliwal11}
\bibinfo{author}{{Kasliwal}, M.~M.} \emph{et~al.}
\newblock \bibinfo{title}{{Discovery of a New Photometric Sub-class of Faint
  and Fast Classical Novae}}.
\newblock \emph{\bibinfo{journal}{\apj}} \textbf{\bibinfo{volume}{735}},
  \bibinfo{pages}{94} (\bibinfo{year}{2011}).

\bibitem{Vinko15}
\bibinfo{author}{{Vink{\'o}}, J.} \emph{et~al.}
\newblock \bibinfo{title}{{A Luminous, Fast Rising UV-transient Discovered by
  ROTSE: A Tidal Disruption Event?}}
\newblock \emph{\bibinfo{journal}{\apj}} \textbf{\bibinfo{volume}{798}},
  \bibinfo{pages}{12} (\bibinfo{year}{2015}).

\bibitem{Greiner15}
\bibinfo{author}{{Greiner}, J.} \emph{et~al.}
\newblock \bibinfo{title}{{A very luminous magnetar-powered supernova
  associated with an ultra-long {\ensuremath{\gamma}}-ray burst}}.
\newblock \emph{\bibinfo{journal}{\nat}} \textbf{\bibinfo{volume}{523}},
  \bibinfo{pages}{189--192} (\bibinfo{year}{2015}).

\bibitem{Rest18}
\bibinfo{author}{{Rest}, A.} \emph{et~al.}
\newblock \bibinfo{title}{{A fast-evolving luminous transient discovered by
  K2/Kepler}}.
\newblock \emph{\bibinfo{journal}{Nature Astronomy}}
  \textbf{\bibinfo{volume}{2}}, \bibinfo{pages}{307--311}
  (\bibinfo{year}{2018}).

\bibitem{Ho19}
\bibinfo{author}{{Ho}, A. Y.~Q.} \emph{et~al.}
\newblock \bibinfo{title}{{The Death Throes of a Stripped Massive Star: An
  Eruptive Mass-Loss History Encoded in Pre-Explosion Emission, a Rapidly
  Rising Luminous Transient, and a Broad-Lined Ic Supernova SN2018gep}}.
\newblock \emph{\bibinfo{journal}{arXiv e-prints}}
  \bibinfo{pages}{arXiv:1904.11009} (\bibinfo{year}{2019}).

\bibitem{Drout14}
\bibinfo{author}{{Drout}, M.~R.} \emph{et~al.}
\newblock \bibinfo{title}{{Rapidly Evolving and Luminous Transients from
  Pan-STARRS1}}.
\newblock \emph{\bibinfo{journal}{\apj}} \textbf{\bibinfo{volume}{794}},
  \bibinfo{pages}{23} (\bibinfo{year}{2014}).

\bibitem{Taddia15_sdss}
\bibinfo{author}{{Taddia}, F.} \emph{et~al.}
\newblock \bibinfo{title}{{Early-time light curves of Type Ib/c supernovae from
  the SDSS-II Supernova Survey}}.
\newblock \emph{\bibinfo{journal}{\aap}} \textbf{\bibinfo{volume}{574}},
  \bibinfo{pages}{A60} (\bibinfo{year}{2015}).

\bibitem{Gall15}
\bibinfo{author}{{Gall}, E.~E.~E.} \emph{et~al.}
\newblock \bibinfo{title}{{A comparative study of Type II-P and II-L supernova
  rise times as exemplified by the case of LSQ13cuw}}.
\newblock \emph{\bibinfo{journal}{\aap}} \textbf{\bibinfo{volume}{582}},
  \bibinfo{pages}{A3} (\bibinfo{year}{2015}).

\bibitem{Taddia16a}
\bibinfo{author}{{Taddia}, F.} \emph{et~al.}
\newblock \bibinfo{title}{{Long-rising Type II supernovae from Palomar
  Transient Factory and Caltech Core-Collapse Project}}.
\newblock \emph{\bibinfo{journal}{\aap}} \textbf{\bibinfo{volume}{588}},
  \bibinfo{pages}{A5} (\bibinfo{year}{2016}).

\bibitem{Arcavi16}
\bibinfo{author}{{Arcavi}, I.} \emph{et~al.}
\newblock \bibinfo{title}{{Rapidly Rising Transients in the
  Supernova{\textemdash}Superluminous Supernova Gap}}.
\newblock \emph{\bibinfo{journal}{\apj}} \textbf{\bibinfo{volume}{819}},
  \bibinfo{pages}{35} (\bibinfo{year}{2016}).

\bibitem{Taddia19}
\bibinfo{author}{{Taddia}, F.} \emph{et~al.}
\newblock \bibinfo{title}{{Analysis of broad-lined Type Ic supernovae from the
  (intermediate) Palomar Transient Factory}}.
\newblock \emph{\bibinfo{journal}{\aap}} \textbf{\bibinfo{volume}{621}},
  \bibinfo{pages}{A71} (\bibinfo{year}{2019}).

\bibitem{Pursiainen18}
\bibinfo{author}{{Pursiainen}, M.} \emph{et~al.}
\newblock \bibinfo{title}{{Rapidly evolving transients in the Dark Energy
  Survey}}.
\newblock \emph{\bibinfo{journal}{\mnras}} \textbf{\bibinfo{volume}{481}},
  \bibinfo{pages}{894--917} (\bibinfo{year}{2018}).

\bibitem{Drout11}
\bibinfo{author}{{Drout}, M.~R.} \emph{et~al.}
\newblock \bibinfo{title}{{The First Systematic Study of Type Ibc Supernova
  Multi-band Light Curves}}.
\newblock \emph{\bibinfo{journal}{\apj}} \textbf{\bibinfo{volume}{741}},
  \bibinfo{pages}{97--117} (\bibinfo{year}{2011}).

\bibitem{Bianco14}
\bibinfo{author}{{Bianco}, F.~B.} \emph{et~al.}
\newblock \bibinfo{title}{{Multi-color Optical and Near-infrared Light Curves
  of 64 Stripped-envelope Core-Collapse Supernovae}}.
\newblock \emph{\bibinfo{journal}{\apjs}} \textbf{\bibinfo{volume}{213}},
  \bibinfo{pages}{19} (\bibinfo{year}{2014}).

\bibitem{Fremling18}
\bibinfo{author}{{Fremling}, C.} \emph{et~al.}
\newblock \bibinfo{title}{{Oxygen and helium in stripped-envelope supernovae}}.
\newblock \emph{\bibinfo{journal}{\aap}} \textbf{\bibinfo{volume}{618}},
  \bibinfo{pages}{A37} (\bibinfo{year}{2018}).

\bibitem{Shivvers19}
\bibinfo{author}{{Shivvers}, I.} \emph{et~al.}
\newblock \bibinfo{title}{{The Berkeley sample of stripped-envelope
  supernovae}}.
\newblock \emph{\bibinfo{journal}{\mnras}} \textbf{\bibinfo{volume}{482}},
  \bibinfo{pages}{1545--1556} (\bibinfo{year}{2019}).

\bibitem{Prentice19}
\bibinfo{author}{{Prentice}, S.~J.} \emph{et~al.}
\newblock \bibinfo{title}{{Investigating the properties of stripped-envelope
  supernovae; what are the implications for their progenitors?}}
\newblock \emph{\bibinfo{journal}{\mnras}} \textbf{\bibinfo{volume}{485}},
  \bibinfo{pages}{1559--1578} (\bibinfo{year}{2019}).

\bibitem{Lyman16}
\bibinfo{author}{{Lyman}, J.~D.} \emph{et~al.}
\newblock \bibinfo{title}{{Bolometric light curves and explosion parameters of
  38 stripped-envelope core-collapse supernovae}}.
\newblock \emph{\bibinfo{journal}{\mnras}} \textbf{\bibinfo{volume}{457}},
  \bibinfo{pages}{328--350} (\bibinfo{year}{2016}).

\bibitem{Graur17a}
\bibinfo{author}{{Graur}, O.} \emph{et~al.}
\newblock \bibinfo{title}{{LOSS Revisited. I. Unraveling Correlations Between
  Supernova Rates and Galaxy Properties, as Measured in a Reanalysis of the
  Lick Observatory Supernova Search}}.
\newblock \emph{\bibinfo{journal}{\apj}} \textbf{\bibinfo{volume}{837}},
  \bibinfo{pages}{120} (\bibinfo{year}{2017}).

\bibitem{Kerzendorf19}
\bibinfo{author}{{Kerzendorf}, W.~E.} \emph{et~al.}
\newblock \bibinfo{title}{{No surviving non-compact stellar companion to
  Cassiopeia A}}.
\newblock \emph{\bibinfo{journal}{\aap}} \textbf{\bibinfo{volume}{623}},
  \bibinfo{pages}{A34} (\bibinfo{year}{2019}).

\bibitem{Krause08}
\bibinfo{author}{{Krause}, O.} \emph{et~al.}
\newblock \bibinfo{title}{{The Cassiopeia A Supernova Was of Type IIb}}.
\newblock \emph{\bibinfo{journal}{Science}} \textbf{\bibinfo{volume}{320}},
  \bibinfo{pages}{1195} (\bibinfo{year}{2008}).

\bibitem{Rest08}
\bibinfo{author}{{Rest}, A.} \emph{et~al.}
\newblock \bibinfo{title}{{Scattered-Light Echoes from the Historical Galactic
  Supernovae Cassiopeia A and Tycho (SN 1572)}}.
\newblock \emph{\bibinfo{journal}{\apjl}} \textbf{\bibinfo{volume}{681}},
  \bibinfo{pages}{L81} (\bibinfo{year}{2008}).

\bibitem{Arnett1982}
\bibinfo{author}{{Arnett}, W.~D.}
\newblock \bibinfo{title}{{Type I supernovae. I - Analytic solutions for the
  early part of the light curve}}.
\newblock \emph{\bibinfo{journal}{\apj}} \textbf{\bibinfo{volume}{253}},
  \bibinfo{pages}{785--797} (\bibinfo{year}{1982}).

\bibitem{Modjaz06}
\bibinfo{author}{{Modjaz}, M.} \emph{et~al.}
\newblock \bibinfo{title}{{Early-Time Photometry and Spectroscopy of the Fast
  Evolving SN 2006aj Associated with GRB 060218}}.
\newblock \emph{\bibinfo{journal}{\apj}} \textbf{\bibinfo{volume}{645}},
  \bibinfo{pages}{L21--L24} (\bibinfo{year}{2006}).

\bibitem{Brown14}
\bibinfo{author}{{Brown}, P.~J.}, \bibinfo{author}{{Breeveld}, A.~A.},
  \bibinfo{author}{{Holland}, S.}, \bibinfo{author}{{Kuin}, P.} \&
  \bibinfo{author}{{Pritchard}, T.}
\newblock \bibinfo{title}{{SOUSA: the Swift Optical/Ultraviolet Supernova
  Archive}}.
\newblock \emph{\bibinfo{journal}{\apss}} \textbf{\bibinfo{volume}{354}},
  \bibinfo{pages}{89--96} (\bibinfo{year}{2014}).

\bibitem{Mazzali08}
\bibinfo{author}{{Mazzali}, P.~A.} \emph{et~al.}
\newblock \bibinfo{title}{{The Metamorphosis of Supernova SN 2008D/XRF 080109:
  A Link Between Supernovae and GRBs/Hypernovae}}.
\newblock \emph{\bibinfo{journal}{Science}} \textbf{\bibinfo{volume}{321}},
  \bibinfo{pages}{1185} (\bibinfo{year}{2008}).

\bibitem{Modjaz09}
\bibinfo{author}{{Modjaz}, M.} \emph{et~al.}
\newblock \bibinfo{title}{{From Shock Breakout to Peak and Beyond: Extensive
  Panchromatic Observations of the Type Ib Supernova 2008D Associated with
  Swift X-ray Transient 080109}}.
\newblock \emph{\bibinfo{journal}{\apj}} \textbf{\bibinfo{volume}{702}},
  \bibinfo{pages}{226--248} (\bibinfo{year}{2009}).

\bibitem{Arcavi11}
\bibinfo{author}{{Arcavi}, I.} \emph{et~al.}
\newblock \bibinfo{title}{{SN 2011dh: Discovery of a Type IIb Supernova from a
  Compact Progenitor in the Nearby Galaxy M51}}.
\newblock \emph{\bibinfo{journal}{\apjl}} \textbf{\bibinfo{volume}{742}},
  \bibinfo{pages}{L18} (\bibinfo{year}{2011}).

\bibitem{Kumar13}
\bibinfo{author}{{Kumar}, B.} \emph{et~al.}
\newblock \bibinfo{title}{{Light curve and spectral evolution of the Type IIb
  supernova 2011fu}}.
\newblock \emph{\bibinfo{journal}{\mnras}} \textbf{\bibinfo{volume}{431}},
  \bibinfo{pages}{308--321} (\bibinfo{year}{2013}).

\bibitem{Morales-Garoffolo14}
\bibinfo{author}{{Morales-Garoffolo}, A.} \emph{et~al.}
\newblock \bibinfo{title}{{SN 2013df, a double-peaked IIb supernova from a
  compact progenitor and an extended H envelope}}.
\newblock \emph{\bibinfo{journal}{\mnras}} \textbf{\bibinfo{volume}{445}},
  \bibinfo{pages}{1647--1662} (\bibinfo{year}{2014}).

\bibitem{Arcavi17a}
\bibinfo{author}{{Arcavi}, I.} \emph{et~al.}
\newblock \bibinfo{title}{{Constraints on the Progenitor of SN 2016gkg from Its
  Shock-cooling Light Curve}}.
\newblock \emph{\bibinfo{journal}{\apj}} \textbf{\bibinfo{volume}{837}},
  \bibinfo{pages}{L2} (\bibinfo{year}{2017}).

\bibitem{Bersten18}
\bibinfo{author}{{Bersten}, M.~C.} \emph{et~al.}
\newblock \bibinfo{title}{{A surge of light at the birth of a supernova}}.
\newblock \emph{\bibinfo{journal}{\nat}} \textbf{\bibinfo{volume}{554}},
  \bibinfo{pages}{497--499} (\bibinfo{year}{2018}).

\bibitem{Izzo19}
\bibinfo{author}{{Izzo}, L.} \emph{et~al.}
\newblock \bibinfo{title}{{Signatures of a jet cocoon in early spectra of a
  supernova associated with a {$\gamma$}-ray burst}}.
\newblock \emph{\bibinfo{journal}{\nat}} \textbf{\bibinfo{volume}{565}},
  \bibinfo{pages}{324--327} (\bibinfo{year}{2019}).

\bibitem{Garnavich16}
\bibinfo{author}{{Garnavich}, P.~M.} \emph{et~al.}
\newblock \bibinfo{title}{{Shock Breakout and Early Light Curves of Type II-P
  Supernovae Observed with Kepler}}.
\newblock \emph{\bibinfo{journal}{\apj}} \textbf{\bibinfo{volume}{820}},
  \bibinfo{pages}{23} (\bibinfo{year}{2016}).

\bibitem{Colgate1969}
\bibinfo{author}{{Colgate}, S.~A.} \& \bibinfo{author}{{McKee}, C.}
\newblock \bibinfo{title}{{Early Supernova Luminosity}}.
\newblock \emph{\bibinfo{journal}{\apj}} \textbf{\bibinfo{volume}{157}},
  \bibinfo{pages}{623} (\bibinfo{year}{1969}).

\bibitem{Weaver1976}
\bibinfo{author}{{Weaver}, T.~A.}
\newblock \bibinfo{title}{{The structure of supernova shock waves}}.
\newblock \emph{\bibinfo{journal}{\apjs}} \textbf{\bibinfo{volume}{32}},
  \bibinfo{pages}{233--282} (\bibinfo{year}{1976}).

\bibitem{Klein1978}
\bibinfo{author}{{Klein}, R.~I.} \& \bibinfo{author}{{Chevalier}, R.~A.}
\newblock \bibinfo{title}{{X-ray bursts from Type II supernovae}}.
\newblock \emph{\bibinfo{journal}{\apjl}} \textbf{\bibinfo{volume}{223}},
  \bibinfo{pages}{L109--L112} (\bibinfo{year}{1978}).

\bibitem{Falk1978}
\bibinfo{author}{{Falk}, S.~W.}
\newblock \bibinfo{title}{{Shock steepening and prompt thermal emission in
  supernovae}}.
\newblock \emph{\bibinfo{journal}{\apjl}} \textbf{\bibinfo{volume}{225}},
  \bibinfo{pages}{L133--L136} (\bibinfo{year}{1978}).

\bibitem{Matzner99}
\bibinfo{author}{{Matzner}, C.~D.} \& \bibinfo{author}{{McKee}, C.~F.}
\newblock \bibinfo{title}{{The Expulsion of Stellar Envelopes in Core-Collapse
  Supernovae}}.
\newblock \emph{\bibinfo{journal}{\apj}} \textbf{\bibinfo{volume}{510}},
  \bibinfo{pages}{379--403} (\bibinfo{year}{1999}).

\bibitem{Nakar_Sari_2010}
\bibinfo{author}{{Nakar}, E.} \& \bibinfo{author}{{Sari}, R.}
\newblock \bibinfo{title}{{Early Supernovae Light Curves Following the Shock
  Breakout}}.
\newblock \emph{\bibinfo{journal}{\apj}} \textbf{\bibinfo{volume}{725}},
  \bibinfo{pages}{904--921} (\bibinfo{year}{2010}).

\bibitem{Rabinak_Waxman_2011}
\bibinfo{author}{{Rabinak}, I.} \& \bibinfo{author}{{Waxman}, E.}
\newblock \bibinfo{title}{{The Early UV/Optical Emission from Core-collapse
  Supernovae}}.
\newblock \emph{\bibinfo{journal}{\apj}} \textbf{\bibinfo{volume}{728}},
  \bibinfo{pages}{63} (\bibinfo{year}{2011}).

\bibitem{Waxman2017}
\bibinfo{author}{{Waxman}, E.} \& \bibinfo{author}{{Katz}, B.}
\newblock \emph{\bibinfo{title}{{Shock Breakout Theory}}}, \bibinfo{pages}{967}
  (\bibinfo{year}{2017}).

\bibitem{Tominaga2011}
\bibinfo{author}{{Tominaga}, N.} \emph{et~al.}
\newblock \bibinfo{title}{{Shock Breakout in Type II Plateau Supernovae:
  Prospects for High-Redshift Supernova Surveys}}.
\newblock \emph{\bibinfo{journal}{\apjs}} \textbf{\bibinfo{volume}{193}},
  \bibinfo{pages}{20} (\bibinfo{year}{2011}).

\bibitem{Schawinski2008}
\bibinfo{author}{{Schawinski}, K.} \emph{et~al.}
\newblock \bibinfo{title}{{Supernova Shock Breakout from a Red Supergiant}}.
\newblock \emph{\bibinfo{journal}{Science}} \textbf{\bibinfo{volume}{321}},
  \bibinfo{pages}{223--226} (\bibinfo{year}{2008}).

\bibitem{Soderberg_2008}
\bibinfo{author}{{Soderberg}, A.~M.} \emph{et~al.}
\newblock \bibinfo{title}{{An extremely luminous X-ray outburst at the birth of
  a supernova}}.
\newblock \emph{\bibinfo{journal}{\nat}} \textbf{\bibinfo{volume}{453}},
  \bibinfo{pages}{469--474} (\bibinfo{year}{2008}).

\bibitem{Gezari2015}
\bibinfo{author}{{Gezari}, S.} \emph{et~al.}
\newblock \bibinfo{title}{{GALEX Detection of Shock Breakout in Type IIP
  Supernova PS1-13arp: Implications for the Progenitor Star Wind}}.
\newblock \emph{\bibinfo{journal}{\apj}} \textbf{\bibinfo{volume}{804}},
  \bibinfo{pages}{28} (\bibinfo{year}{2015}).

\bibitem{Couch11}
\bibinfo{author}{{Couch}, S.~M.}, \bibinfo{author}{{Pooley}, D.},
  \bibinfo{author}{{Wheeler}, J.~C.} \& \bibinfo{author}{{Milosavljevi{\'c}},
  M.}
\newblock \bibinfo{title}{{Aspherical Supernova Shock Breakout and the
  Observations of Supernova 2008D}}.
\newblock \emph{\bibinfo{journal}{\apj}} \textbf{\bibinfo{volume}{727}},
  \bibinfo{pages}{104} (\bibinfo{year}{2011}).

\bibitem{Svirski14}
\bibinfo{author}{{Svirski}, G.} \& \bibinfo{author}{{Nakar}, E.}
\newblock \bibinfo{title}{{SN 2008D: A Wolf-Rayet Explosion Through a Thick
  Wind}}.
\newblock \emph{\bibinfo{journal}{\apjl}} \textbf{\bibinfo{volume}{788}},
  \bibinfo{pages}{L14} (\bibinfo{year}{2014}).

\bibitem{Campana06}
\bibinfo{author}{{Campana}, S.} \emph{et~al.}
\newblock \bibinfo{title}{{The association of GRB 060218 with a supernova and
  the evolution of the shock wave}}.
\newblock \emph{\bibinfo{journal}{\nat}} \textbf{\bibinfo{volume}{442}},
  \bibinfo{pages}{1008--1010} (\bibinfo{year}{2006}).

\bibitem{Nakar12}
\bibinfo{author}{{Nakar}, E.} \& \bibinfo{author}{{Sari}, R.}
\newblock \bibinfo{title}{{Relativistic Shock Breakouts - A Variety of
  Gamma-Ray Flares: From Low-luminosity Gamma-Ray Bursts to Type Ia
  Supernovae}}.
\newblock \emph{\bibinfo{journal}{\apj}} \textbf{\bibinfo{volume}{747}},
  \bibinfo{pages}{88} (\bibinfo{year}{2012}).

\bibitem{Irwin16}
\bibinfo{author}{{Irwin}, C.~M.} \& \bibinfo{author}{{Chevalier}, R.~A.}
\newblock \bibinfo{title}{{Jet or shock breakout? The low-luminosity GRB
  060218}}.
\newblock \emph{\bibinfo{journal}{\mnras}} \textbf{\bibinfo{volume}{460}},
  \bibinfo{pages}{1680--1704} (\bibinfo{year}{2016}).

\bibitem{Garnavich_2016}
\bibinfo{author}{{Garnavich}, P.~M.} \emph{et~al.}
\newblock \bibinfo{title}{{Shock Breakout and Early Light Curves of Type II-P
  Supernovae Observed with Kepler}}.
\newblock \emph{\bibinfo{journal}{\apj}} \textbf{\bibinfo{volume}{820}},
  \bibinfo{pages}{23} (\bibinfo{year}{2016}).

\bibitem{Rubin2017}
\bibinfo{author}{{Rubin}, A.} \& \bibinfo{author}{{Gal-Yam}, A.}
\newblock \bibinfo{title}{{Exploring the Efficacy and Limitations of
  Shock-cooling Models: New Analysis of Type II Supernovae Observed by the
  Kepler Mission}}.
\newblock \emph{\bibinfo{journal}{\apj}} \textbf{\bibinfo{volume}{848}},
  \bibinfo{pages}{8} (\bibinfo{year}{2017}).

\bibitem{Ofek2010}
\bibinfo{author}{{Ofek}, E.~O.} \emph{et~al.}
\newblock \bibinfo{title}{{Supernova PTF 09UJ: A Possible Shock Breakout from a
  Dense Circumstellar Wind}}.
\newblock \emph{\bibinfo{journal}{\apj}} \textbf{\bibinfo{volume}{724}},
  \bibinfo{pages}{1396--1401} (\bibinfo{year}{2010}).

\bibitem{Chevalier2011}
\bibinfo{author}{{Chevalier}, R.~A.} \& \bibinfo{author}{{Irwin}, C.~M.}
\newblock \bibinfo{title}{{Shock Breakout in Dense Mass Loss: Luminous
  Supernovae}}.
\newblock \emph{\bibinfo{journal}{\apjl}} \textbf{\bibinfo{volume}{729}},
  \bibinfo{pages}{L6} (\bibinfo{year}{2011}).

\bibitem{Balberg2011}
\bibinfo{author}{{Balberg}, S.} \& \bibinfo{author}{{Loeb}, A.}
\newblock \bibinfo{title}{{Supernova shock breakout through a wind}}.
\newblock \emph{\bibinfo{journal}{\mnras}} \textbf{\bibinfo{volume}{414}},
  \bibinfo{pages}{1715--1720} (\bibinfo{year}{2011}).

\bibitem{Svirski2012}
\bibinfo{author}{{Svirski}, G.}, \bibinfo{author}{{Nakar}, E.} \&
  \bibinfo{author}{{Sari}, R.}
\newblock \bibinfo{title}{{Optical to X-Ray Supernova Light Curves Following
  Shock Breakout through a Thick Wind}}.
\newblock \emph{\bibinfo{journal}{\apj}} \textbf{\bibinfo{volume}{759}},
  \bibinfo{pages}{108} (\bibinfo{year}{2012}).

\bibitem{Ginzburg2014}
\bibinfo{author}{{Ginzburg}, S.} \& \bibinfo{author}{{Balberg}, S.}
\newblock \bibinfo{title}{{Light Curves from Supernova Shock Breakout through
  an Extended Wind}}.
\newblock \emph{\bibinfo{journal}{\apj}} \textbf{\bibinfo{volume}{780}},
  \bibinfo{pages}{18} (\bibinfo{year}{2014}).

\bibitem{Moriya2017}
\bibinfo{author}{{Moriya}, T.~J.}, \bibinfo{author}{{Yoon}, S.-C.},
  \bibinfo{author}{{Gr{\"a}fener}, G.} \& \bibinfo{author}{{Blinnikov}, S.~I.}
\newblock \bibinfo{title}{{Immediate dense circumstellar environment of
  supernova progenitors caused by wind acceleration: its effect on supernova
  light curves}}.
\newblock \emph{\bibinfo{journal}{\mnras}} \textbf{\bibinfo{volume}{469}},
  \bibinfo{pages}{L108--L112} (\bibinfo{year}{2017}).

\bibitem{Forster2018}
\bibinfo{author}{{Forster}, F.} \emph{et~al.}
\newblock \bibinfo{title}{{The delay of shock breakout due to circumstellar
  material evident in most type II supernovae.}}
\newblock \emph{\bibinfo{journal}{Nature Astronomy}}
  \textbf{\bibinfo{volume}{2}}, \bibinfo{pages}{808--818}
  (\bibinfo{year}{2018}).

\bibitem{Chevalier08}
\bibinfo{author}{{Chevalier}, R.~A.} \& \bibinfo{author}{{Fransson}, C.}
\newblock \bibinfo{title}{{Shock Breakout Emission from a Type Ib/c Supernova:
  XRT 080109/SN 2008D}}.
\newblock \emph{\bibinfo{journal}{\apjl}} \textbf{\bibinfo{volume}{683}},
  \bibinfo{pages}{L135} (\bibinfo{year}{2008}).

\bibitem{Nakar_Piro_2014}
\bibinfo{author}{{Nakar}, E.} \& \bibinfo{author}{{Piro}, A.~L.}
\newblock \bibinfo{title}{{Supernovae with Two Peaks in the Optical Light Curve
  and the Signature of Progenitors with Low-mass Extended Envelopes}}.
\newblock \emph{\bibinfo{journal}{\apj}} \textbf{\bibinfo{volume}{788}},
  \bibinfo{pages}{193} (\bibinfo{year}{2014}).

\bibitem{Piro_2015}
\bibinfo{author}{{Piro}, A.~L.}
\newblock \bibinfo{title}{{Using Double-peaked Supernova Light Curves to Study
  Extended Material}}.
\newblock \emph{\bibinfo{journal}{\apjl}} \textbf{\bibinfo{volume}{808}},
  \bibinfo{pages}{L51} (\bibinfo{year}{2015}).

\bibitem{Sapir_Waxman_2017}
\bibinfo{author}{{Sapir}, N.} \& \bibinfo{author}{{Waxman}, E.}
\newblock \bibinfo{title}{{UV/Optical Emission from the Expanding Envelopes of
  Type II Supernovae}}.
\newblock \emph{\bibinfo{journal}{\apj}} \textbf{\bibinfo{volume}{838}},
  \bibinfo{pages}{130} (\bibinfo{year}{2017}).

\bibitem{Stritzinger02}
\bibinfo{author}{{Stritzinger}, M.} \emph{et~al.}
\newblock \bibinfo{title}{{Optical Photometry of the Type Ia Supernova 1999ee
  and the Type Ib/c Supernova 1999ex in IC 5179}}.
\newblock \emph{\bibinfo{journal}{\aj}} \textbf{\bibinfo{volume}{124}},
  \bibinfo{pages}{2100--2117} (\bibinfo{year}{2002}).

\bibitem{Hoflich1993}
\bibinfo{author}{{Hoflich}, P.}, \bibinfo{author}{{Langer}, N.} \&
  \bibinfo{author}{{Duschinger}, M.}
\newblock \bibinfo{title}{{Supernova 1993J - Explosion of a Massive Cool
  Supergiant with a Small Envelope Mass}}.
\newblock \emph{\bibinfo{journal}{\aap}} \textbf{\bibinfo{volume}{275}},
  \bibinfo{pages}{L29} (\bibinfo{year}{1993}).

\bibitem{Bersten12}
\bibinfo{author}{{Bersten}, M.~C.} \emph{et~al.}
\newblock \bibinfo{title}{{The Type IIb Supernova 2011dh from a Supergiant
  Progenitor}}.
\newblock \emph{\bibinfo{journal}{\apj}} \textbf{\bibinfo{volume}{757}},
  \bibinfo{pages}{31} (\bibinfo{year}{2012}).

\bibitem{Benvenuto2013}
\bibinfo{author}{{Benvenuto}, O.~G.}, \bibinfo{author}{{Bersten}, M.~C.} \&
  \bibinfo{author}{{Nomoto}, K.}
\newblock \bibinfo{title}{{A Binary Progenitor for the Type IIb Supernova
  2011dh in M51}}.
\newblock \emph{\bibinfo{journal}{\apj}} \textbf{\bibinfo{volume}{762}},
  \bibinfo{pages}{74} (\bibinfo{year}{2013}).

\bibitem{GalYam14}
\bibinfo{author}{{Gal-Yam}, A.} \emph{et~al.}
\newblock \bibinfo{title}{{A Wolf-Rayet-like progenitor of SN 2013cu from
  spectral observations of a stellar wind}}.
\newblock \emph{\bibinfo{journal}{\nat}} \textbf{\bibinfo{volume}{509}},
  \bibinfo{pages}{471--474} (\bibinfo{year}{2014}).

\bibitem{Shivvers15}
\bibinfo{author}{{Shivvers}, I.} \emph{et~al.}
\newblock \bibinfo{title}{{Early Emission from the Type IIn Supernova 1998S at
  High Resolution}}.
\newblock \emph{\bibinfo{journal}{\apj}} \textbf{\bibinfo{volume}{806}},
  \bibinfo{pages}{213} (\bibinfo{year}{2015}).

\bibitem{Yaron17}
\bibinfo{author}{{Yaron}, O.} \emph{et~al.}
\newblock \bibinfo{title}{{Confined dense circumstellar material surrounding a
  regular type II supernova}}.
\newblock \emph{\bibinfo{journal}{Nature Physics}}
  \textbf{\bibinfo{volume}{13}}, \bibinfo{pages}{510--517}
  (\bibinfo{year}{2017}).

\bibitem{Morozova2017}
\bibinfo{author}{{Morozova}, V.}, \bibinfo{author}{{Piro}, A.~L.} \&
  \bibinfo{author}{{Valenti}, S.}
\newblock \bibinfo{title}{{Unifying Type II Supernova Light Curves with Dense
  Circumstellar Material}}.
\newblock \emph{\bibinfo{journal}{\apj}} \textbf{\bibinfo{volume}{838}},
  \bibinfo{pages}{28} (\bibinfo{year}{2017}).

\bibitem{Morozova2018}
\bibinfo{author}{{Morozova}, V.}, \bibinfo{author}{{Piro}, A.~L.} \&
  \bibinfo{author}{{Valenti}, S.}
\newblock \bibinfo{title}{{Measuring the Progenitor Masses and Dense
  Circumstellar Material of Type II Supernovae}}.
\newblock \emph{\bibinfo{journal}{\apj}} \textbf{\bibinfo{volume}{858}},
  \bibinfo{pages}{15} (\bibinfo{year}{2018}).

\bibitem{Khazov2016}
\bibinfo{author}{{Khazov}, D.} \emph{et~al.}
\newblock \bibinfo{title}{{Flash Spectroscopy: Emission Lines from the Ionized
  Circumstellar Material around <10-day-old Type II Supernovae}}.
\newblock \emph{\bibinfo{journal}{\apj}} \textbf{\bibinfo{volume}{818}},
  \bibinfo{pages}{3} (\bibinfo{year}{2016}).

\bibitem{ZTF2019}
\bibinfo{author}{{Bellm}, E.~C.} \emph{et~al.}
\newblock \bibinfo{title}{{The Zwicky Transient Facility: System Overview,
  Performance, and First Results}}.
\newblock \emph{\bibinfo{journal}{\pasp}} \textbf{\bibinfo{volume}{131}},
  \bibinfo{pages}{018002} (\bibinfo{year}{2019}).

\bibitem{LSST2008}
\bibinfo{author}{{Ivezic}, Z.} \emph{et~al.}
\newblock \bibinfo{title}{{Large Synoptic Survey Telescope: From Science
  Drivers To Reference Design}}.
\newblock \emph{\bibinfo{journal}{Serbian Astronomical Journal}}
  \textbf{\bibinfo{volume}{176}}, \bibinfo{pages}{1--13}
  (\bibinfo{year}{2008}).

\bibitem{DLT2018}
\bibinfo{author}{{Tartaglia}, L.} \emph{et~al.}
\newblock \bibinfo{title}{{The Early Detection and Follow-up of the Highly
  Obscured Type II Supernova 2016ija/DLT16am}}.
\newblock \emph{\bibinfo{journal}{\apj}} \textbf{\bibinfo{volume}{853}},
  \bibinfo{pages}{62} (\bibinfo{year}{2018}).

\bibitem{Kochanek2017}
\bibinfo{author}{{Kochanek}, C.~S.} \emph{et~al.}
\newblock \bibinfo{title}{{The All-Sky Automated Survey for Supernovae
  (ASAS-SN) Light Curve Server v1.0}}.
\newblock \emph{\bibinfo{journal}{\pasp}} \textbf{\bibinfo{volume}{129}},
  \bibinfo{pages}{104502} (\bibinfo{year}{2017}).

\bibitem{SEDM2018}
\bibinfo{author}{{Blagorodnova}, N.} \emph{et~al.}
\newblock \bibinfo{title}{{The SED Machine: A Robotic Spectrograph for Fast
  Transient Classification}}.
\newblock \emph{\bibinfo{journal}{Publications of the Astronomical Society of
  the Pacific}} \textbf{\bibinfo{volume}{130}}, \bibinfo{pages}{035003}
  (\bibinfo{year}{2018}).

\bibitem{LCOGT2013}
\bibinfo{author}{{Brown}, T.~M.} \emph{et~al.}
\newblock \bibinfo{title}{{Las Cumbres Observatory Global Telescope Network}}.
\newblock \emph{\bibinfo{journal}{\pasp}} \textbf{\bibinfo{volume}{125}},
  \bibinfo{pages}{1031} (\bibinfo{year}{2013}).

\bibitem{Swift2004}
\bibinfo{author}{{Gehrels}, N.} \emph{et~al.}
\newblock \bibinfo{title}{{The Swift Gamma-Ray Burst Mission}}.
\newblock \emph{\bibinfo{journal}{\apj}} \textbf{\bibinfo{volume}{611}},
  \bibinfo{pages}{1005--1020} (\bibinfo{year}{2004}).

\bibitem{ULTRASAT2014}
\bibinfo{author}{{Sagiv}, I.} \emph{et~al.}
\newblock \bibinfo{title}{{Science with a Wide-field UV Transient Explorer}}.
\newblock \emph{\bibinfo{journal}{\aj}} \textbf{\bibinfo{volume}{147}},
  \bibinfo{pages}{79} (\bibinfo{year}{2014}).

\bibitem{CUTIE2017}
\bibinfo{author}{{Cenko}, S.~B.} \emph{et~al.}
\newblock \bibinfo{title}{{CUTIE: Cubesat Ultraviolet Transient Imaging
  Experiment}}.
\newblock In \emph{\bibinfo{booktitle}{American Astronomical Society Meeting
  Abstracts \#229}}, vol. \bibinfo{volume}{229} of
  \emph{\bibinfo{series}{American Astronomical Society Meeting Abstracts}},
  \bibinfo{pages}{328.04} (\bibinfo{year}{2017}).

\bibitem{WISEREP2012}
\bibinfo{author}{{Yaron}, O.} \& \bibinfo{author}{{Gal-Yam}, A.}
\newblock \bibinfo{title}{{WISeREP - An Interactive Supernova Data
  Repository}}.
\newblock \emph{\bibinfo{journal}{\pasp}} \textbf{\bibinfo{volume}{124}},
  \bibinfo{pages}{668} (\bibinfo{year}{2012}).

\bibitem{OSC2017}
\bibinfo{author}{{Guillochon}, J.}, \bibinfo{author}{{Parrent}, J.},
  \bibinfo{author}{{Kelley}, L.~Z.} \& \bibinfo{author}{{Margutti}, R.}
\newblock \bibinfo{title}{{An Open Catalog for Supernova Data}}.
\newblock \emph{\bibinfo{journal}{\apj}} \textbf{\bibinfo{volume}{835}},
  \bibinfo{pages}{64} (\bibinfo{year}{2017}).

\bibitem{TOM2018}
\bibinfo{author}{{Street}, R.~A.}, \bibinfo{author}{{Bowman}, M.},
  \bibinfo{author}{{Saunders}, E.~S.} \& \bibinfo{author}{{Boroson}, T.}
\newblock \bibinfo{title}{{General-purpose software for managing astronomical
  observing programs in the LSST era}}.
\newblock In \emph{\bibinfo{booktitle}{Software and Cyberinfrastructure for
  Astronomy V}}, vol. \bibinfo{volume}{10707} of \emph{\bibinfo{series}{Society
  of Photo-Optical Instrumentation Engineers (SPIE) Conference Series}},
  \bibinfo{pages}{1070711} (\bibinfo{year}{2018}).
\newblock \eprint{1806.09557}.

\end{thebibliography}

\begin{addendum}
 \item[Correspondence] Correspondence and requests for materials
should be addressed to M.M..~(email: mmodjaz@nyu.edu).

\item  We thank E. Nakar, T. Piro, F. Taddia, S. Valenti, and E. Waxman for valuable comments. MM is supported by the NSF CAREER award AST-1352405, by the NSF award AST-1413260 and by a Faculty Fellowship from the Humboldt Foundation. CPG acknowledges support from EU/FP7-ERC grant no. [615929]. IA acknowledges support from the Israel Science Foundation (grant number 2108/18).
 
 \item[Competing Interests] The authors declare that they have no
competing financial interests.

\end{addendum}

\end{document}